# Assessing the Role of Intersection Proximity in Pedestrian Crashes: Insights from Data Mining Approach

**Ahmed Hossain, Ph.D.**
*(Corresponding Author)*
Traffic Safety Analysis Engineer, Multimodal Planning Division (MPD)
Arizona Department of Transportation (ADOT)
206 S 17th Ave, Phoenix, Arizona, 85007
Email: ahossain2@azdot.gov
ORCID: 0000-0003-1566-3993

**Xiaoduan Sun, Ph.D., P.E.**
Professor
Department of Civil Engineering
University of Louisiana at Lafayette, Lafayette, Louisiana, 70503
Email: xsun@louisiana.edu
ORCID: 0000-0001-7282-1340

## ABSTRACT

Although intersections are the most complex parts of the roadway network, pedestrian crashes at non-intersection locations are disproportionately frequent, highlighting a serious traffic safety concern. This study investigates non-intersection crashes involving pedestrians using a crash database (2017-2021) collected from Louisiana State. As the risk of pedestrian crashes tends to vary with distance from the intersection, the research team utilized a unique framework, 'distance to intersection' to capture the differences in crash patterns at non-intersection locations. The study identified that around 50% of non-intersection pedestrian crashes occurred within 198 ft. of the intersection. In the next step, the collected 3,135 pedestrian crashes at non-intersection locations during the study period were subdivided into three zones: D1 zone designates crashes occurring within 150 ft. of an intersection (1,277 crashes), D2 zone designates crashes occurring within 151 ft. to 435 ft. of an intersection (1,060 crashes) and D3 zone designates crashes occurring at 435 ft. or higher from an intersection (798 crashes). To explore the complex interaction of multiple factors, an intuitive data mining technique, Association Rules Mining (ARM), was used. A total of the top 60 interesting association rules (20 for each zone) were identified by the algorithm (based on lift and support measures). In addition, a total of 124 rules were explored based on the Lift Increase Criterion (LIC) measure. The findings of this research provide critical insights into pedestrian crash involvement at non-intersection locations and the variation in crash patterns according to the 'distance to intersection'. Based on the findings, some of the targeted problem-specific countermeasures are also recommended to address the crash patterns at non-intersection locations. The transportation officials can utilize the findings and suggested recommendations for reducing crash risk and improving road safety for pedestrians.

# 1. INTRODUCTION

A total of 7,388 pedestrians were killed in traffic crashes in the U.S. in 2021, which is a 12.5% increase compared to 2020 (*Fatality Facts*, 2021). Most of these pedestrian fatalities occurred at non-intersection locations (i.e., any locations away from intersections). Since pedestrians are less anticipated by drivers at non-intersection locations, their involvement in crashes frequently results in severe injuries and fatalities. The Fatality Analysis and Reporting System (FARS) data suggests that a significant portion of pedestrian fatalities occurred at non-intersection locations in the U.S. during the 2017-2021 period (2017 = 72.9%, 2018 = 73.8%, 2019 = 73.1%, 2020 = 75.8%, 2021 = 76.8%). Mirroring the US national-level trend, pedestrian safety at non-intersection locations is also identified as a major concern in the state of Louisiana, with a substantially greater percentage of fatalities occurring at non-intersections compared to intersections (more details in **Figure 1**). Overall, pedestrian safety has emerged as a significant concern in Louisiana state, which ranks second in the nation with 3.98 pedestrian fatalities per 100,000 population and remains well above the US national average (2.23 pedestrian fatalities per 100,000 population). The Federal Highway Administration (FHWA) designated Louisiana as a priority state addressing pedestrian safety as one of the 'focus area' in 2021. The disproportionately high number of pedestrian fatalities occurring in places other than intersections establishes this as a serious traffic safety hazard at the national and state levels. The current situation warrants immediate actions to conduct crash data investigations focusing on non-intersection locations and develop problem-specific countermeasures to reduce pedestrian crashes.

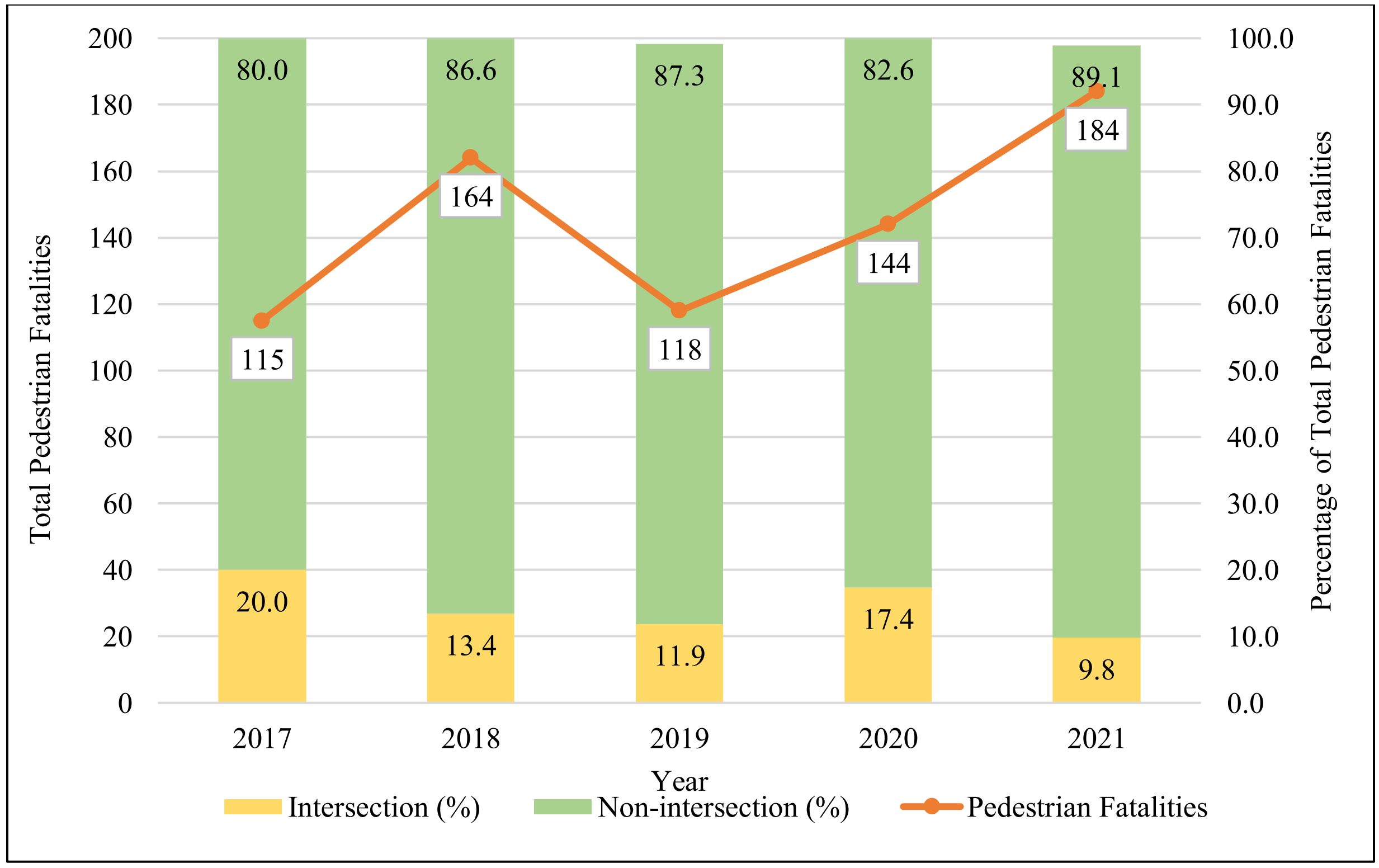

**Figure 1. Pedestrian Fatal Crash Statistics in Louisiana State (2017-2021) [Source: FARS]**

In an ideal case scenario, pedestrians are expected to cross the road at marked crosswalks at the intersections or at segment locations (if a marked crosswalk is available). However, previous investigation on pedestrian crossing behavior suggests that pedestrians usually cross the road at undesignated locations (e.g., no crosswalk) away from intersections (Thakur & Biswas, 2019; L. Thomas et al., 2017; Zuniga-Garcia et al., 2022). The following are the few reasons why pedestrians may choose to cross at midblock (or non-intersection locations) instead of at intersections (Sandt & Zegeer, 2006a) – a) crossing at non-intersection locations may represent the shortest and most convenient path for pedestrian trips; b) many pedestrians are reluctant (or unable) to cross at the nearest intersection, especially if doing so requires walking long distances to the intersection or if there is no obvious benefit to doing so (absence of traffic signals at the intersection) and c) crossing in mid-block sites may be deemed safer or simpler than crossing at an intersection because pedestrians are required to interact with turning (right/left) vehicles at the intersection. Such factors increase the odds of pedestrian crashes at non-intersection locations and are identified as a high-risk scenario as the driver does not expect pedestrians at such locations and is less likely to yield due to higher vehicle speed and less reaction time.

Pedestrian crashes occurring at a close distance to an intersection may not exhibit the same characteristics as those crashes that occur far away from an intersection. This could be due to several traffic (e.g., volume, flow type) and built environment (e.g., land use, roadway design factors) (Dai & Jaworski, 2016; S. Lee et al., 2020). Intersections are points where different streams of traffic converge. Close to an intersection, there is often a higher density of vehicles, frequent lane changes, and turning movements. The complexity of traffic flow and associated traffic volume is completely different at a far distance from the intersection. From a driver's perspective, drivers may behave differently when approaching an intersection. Additionally, the presence of buildings, structures, or other obstacles close to an intersection can limit drivers' sight distance. All these factors (single or a complex interaction of multiple factors) are expected to uniquely affect pedestrian crashes according to the 'distance to intersection' and further research is required in this context

Investigating pedestrian crash patterns based on the 'distance to intersection' offers a novel perspective for addressing pedestrian safety at non-intersection locations and formulating targeted countermeasures. Recognizing this critical dimension, the current study leverages Louisiana crash data (2017-2021) to explore the influence of various contributing factors on pedestrian crashes at non-intersection sites. By adopting 'distance to intersection' as a key analytical metric, this research aims to uncover insights into pedestrian crash dynamics. The study's findings are anticipated to significantly enhance the understanding of safety challenges at non-intersection locations, offering evidence-based recommendations for the development of effective countermeasures.

## 2. LITERATURE REVIEW

Despite non-intersection pedestrian crashes being a frequent occurrence on roadways, research on these incidents remains limited. Previous investigations have predominantly focused on intersection crashes or treated non-intersection crashes as part of broader, aggregated analyses. A recent investigation applied a mixed logit model to examine pedestrian injury severities at intersection and non-intersection locations (Qiu & Fan, 2022). According to the findings, male drivers, alcohol involvement, elderly pedestrians (65+), trucks, and higher posted speed limit significantly increase the likelihood of pedestrian serious injury severities in both intersection and non-intersection locations, while traffic control, adverse weather conditions, and day-of-week have impacts at non-intersection locations. Another similar study used a t-test to compare midblock and intersection crashes involving pedestrians using crash data collected from Kentucky, Florida, and North Carolina (Sandt & Zegeer, 2006b). The study identified significantly higher pedestrian crashes in some specific scenarios including two-lane roadways, young male pedestrians, residential land use patterns, and rural areas.

Another study evaluated crossing-related pedestrian crashes at undesignated midblock locations using a wide range of data analysis techniques including frequency-based descriptive analysis, Chi-squared test, Ordered Probit Model, and Random Forest model (Kwayu et al., 2022). The major findings from this research were: a) a pedestrian was involved in a crash while crossing at an undesignated area about 130 feet (on average) from the nearest intersection, b) the factors contributing to pedestrian crashes at undesignated midblock locations were dark lighting conditions, pedestrian age, dark clothing, and traffic volume. Another study analyzed pedestrian crossing volumes at 64 midblock locations in Texas (Houston and San Antonio) (Reddy Geedipally, 2022). The study also developed Safety Performance Functions (SPFs) that identified traffic volume and pedestrian crossing volume as the most influencing factor contributing to pedestrian crashes.

Some of the previous investigations focused on other spatial, geometric, and traffic control features that affect pedestrian crashes at non-intersection locations. For example, a previous research utilized an ordered probit model to investigate pedestrian crashes in rural Connecticut region based on two considerations: a) pedestrians were attempting to cross two-lane highways, and b) the crash location was not controlled by stop signs or signals (Zajac & Ivan, 2003). Some of the identified factors that influenced pedestrian injury severity were roadway width, type of vehicle, driver or pedestrian impairment, and elderly pedestrians (65 years or older). Using latent class clustering and ordered probit model, a previous study investigated the differences in crash patterns between intersection and non-intersections in the United Kingdom (Salehian et al., 2023). The study reported the day of the week, pedestrian location in a refuge island, and minor roads as a significant factor contributing to injury severity at the rural non-intersection location. Another study investigated the differences in pedestrian crash patterns between urban and rural locations in Louisiana using a multinomial logit model (Sun & Sun, 2020a). The study reported that the likelihood of fatal or severe crash involvement at rural non-intersection locations is 128% higher than at intersections. A different result was reported by another study comparing at-fault pedestrian crashes on rural and urban roadways in Alabama using a random parameter logit model (S. Islam & Jones, 2014a). According to the findings, at-fault pedestrian crashes at rural non-intersection locations have increased the likelihood of no injuries (no injury = 131.2%, minor injury = -6.4%, major injury = -12.4%). Overall, the outcome of pedestrian crashes in rural settings can be associated with the availability of Emergency Medical Services (EMS) which can be limited in such areas (Mueller et al., 1988).

Some of the previous studies focused specifically on the presence of crosswalks and its relationship to pedestrian crash involvement at intersection or non-intersection locations. A previous study analyzed pedestrian injury severity at signalized and non-signalized locations and reported a 1.36% reduction in severity at unsignalized intersections with the presence of a standard crosswalk (Haleem et al., 2015). A similar result was obtained by another research using the Heteroskedastic logit model on police-reported crash data (1997-2000) from North Carolina (Kim et al., 2008). According to the study, a decreased likelihood of fatal, possible, or no injury was observed while pedestrians were crossing roads at designated crosswalks. Another study conducted in Hawaii reported more pedestrian crashes in non-crosswalk locations (61.8%) compared to crosswalk locations (38.2%). Overall, the presence of crosswalks appears to be effective in reducing the likelihood of severe injuries as drivers are alerted to the potential presence of pedestrians at such locations.

From a general context, most of the previous research on pedestrian safety utilized crash location (intersection, non-intersection) as a variable in the model to address pedestrian risk factors at non-intersection locations. **Table 1** provides an overview of these investigations, including the methodology used, study area, and major findings.

**Table 1. Key Findings from Selected Studies**

| Study | Methods; Location | Key findings related to non-intersection/midblock pedestrian crashes |
| --- | --- | --- |
| (Pande et al., 2010) | Classification Trees; US Route 19, Florida | • Alcohol-involved pedestrians are significantly more likely to be involved in crashes on Friday and Saturday nights.<br>• Road segments with higher ADT (>54,125) and wide medians (24 ft. or 28 ft.) increased the likelihood of pedestrian crashes.<br>• The presence of sidewalks has no substantial impact on the likelihood of a pedestrian crash, but parking along the roadside does. |
| (Ferenchak & Abadi, 2021) | ANOVA and two-sample t-test, linear regression, and Breusch-Godfrey test; US | • Infrastructure-related factors have a strong correlation with the rise in fatal pedestrian crashes during nighttime.<br>• Most of the pedestrian fatalities occurred on roadways with posted speed limits of 40–45 mph and unmarked locations that are away from intersections. |
| (Ha & Thill, 2011) | Hot spot spatial analysis and spatial econometric modeling; Buffalo, New York, US | • Young pedestrians socializing or playing in their own residential areas are more likely to be involved in mid-block pedestrian collisions.<br>• Vehicles that are backing off driveways are also responsible for a significant percentage of crashes that occur in the mid-block.<br>• Males are nearly twice as likely as females to be involved in a crash (63% vs. 37%) considering the mid-block location.<br>• Most mid-block collisions occur while vehicles are proceeding straight, rather than making any additional maneuvers like turning, reversing, or parking. |
| (Quistberg et al., 2015) | Multilevel mixed-effects Poisson regression and spatiotemporal approach; Seattle, Washington, USA | • Cul-de-sacs (i.e., street that is closed at one end) and road curves experienced significantly lower accident rates in comparison with mid-blocks.<br>• Nearly 2.5 times fewer collisions occurred at mid-blocks than at intersections with four or five or more lanes. |
| (Bennet & Yiannakoulias, 2015) | Conditional logistic regression; Hamilton, Ontario, Canada | • Mid-block model finds significance for Child Activity using Shortest Distance (CASD) and road segment length, both of which indicate an increased higher risk of collision engaging child pedestrians.<br>• The expected association between traffic volume and collision risk may not hold if low traffic volumes encourage more dangerous mid-block behaviors like jay walking. |
| (Hezaveh & Cherry, 2018) | Binary logistic regression; Tennessee, USA | • The proportion of mid-block collisions was considerably greater for pedestrian walking under the influence (WUI) collisions (69%) than for non-WUI collisions (55%), which worsens at night due to impaired judgments and vision, as well as stepping into the path of a moving vehicle.<br>• Mid-block WUI was also significantly correlated with driver moves, such as the straight maneuver. |
| (Sun et al., 2019) | Latent Class Cluster (LCC) model and Multinomial Logit (MNL) model; Louisiana, USA | • The likelihood of fatal pedestrian crashes increases compared to non-intersection locations due to poor visibility.<br>• Crash fatalities and serious injuries are more likely to occur at speeds of 60 mph or higher when they occur away from intersections in rural areas. |

| Study | Methods; Location | Key findings related to non-intersection/midblock pedestrian crashes |
|---|---|---|
| | | • Pedestrians on low-speed roads in rural, non-intersection locations were 88.2% less likely to sustain fatal or severe injuries compared to those on high-speed zones. |
| (Peng et al., 2020) | SEM Path analysis, multinomial and ordered logit model; USA | • Heavier vehicles (light trucks, buses, and heavy trucks) tend to increase the probability of fatal pedestrian casualties.<br>• When comparing pedestrians aged 25-45, 45-65, and >65, the marginal effects suggest a 27.3%, 31.1%, and 27.1% reduction in the probability of darting or running into the road, respectively.<br>• A pedestrian's likelihood of committing the improper mid-block crossing increases by 5.8%, 8.9%, and 7.3% when crossing a three, four, or five-lane road, respectively.<br>• Pedestrians are 7.1% more likely to be involved in a dart or run into the road when the surface of the pavement is dry, as opposed to when it is wet. |
| (Tanishita et al., 2023) | Bias-reduced Logistic Regression; Japan | • The study finds that the installation of medians raises the probability of fatalities at non-intersection locations, where unlawful conduct is more likely to occur. |

## 2.1 Research Gap and Study Objectives

A critical gap in the current literature has been identified, wherein most studies addressing pedestrian crashes at non-intersection locations rely on gross-level databases and fail to account for how crash characteristics vary based on proximity to intersections. The dynamics of pedestrian crashes are influenced by the complexity of traffic flow, built environment, and behavioral patterns, which differ markedly between crashes occurring near intersections and those further away. This research seeks to address this gap by examining non-intersection pedestrian crashes through the lens of 'distance to the intersection,' aiming to uncover distinct patterns associated with crash occurrence. By providing insights into how pedestrian crashes vary with proximity to intersections, this study offers a more comprehensive understanding of non-intersection crashes. The findings are expected to inform policymakers and safety professionals in the development of targeted countermeasures, enabling more effective interventions tailored to specific roadway contexts.

## 3. METHODS

Association Rules Mining (ARM) is a data mining technique that discovers items (i.e., variable categories) that appear together in a database. Among several algorithms used for ARM, this study utilized the 'Apriori' algorithm due to its advantages over parametric and other non-parametric methods. ARM algorithm has been extensively used in previous safety literature for identifying the association knowledge (i.e., group of risk factors that occur together in the event of a crash) (Xu et al., 2018).

In the context of this research, this study defines an itemset as a set of items that include at least one reported pedestrian crash that occurred at midblock locations. An m-itemset is one that contains m items. For an itemset, $I = \{i_1, i_2, \dots\dots\dots, i_m\}$ of m distinct attributes, consider a crash database, $D = \{t_1, t_2, \dots\dots\dots, t_n\}$ consisting of pedestrian-vehicle crash information such that each crash-risk factor in D is a subset of items contained in I. An association rule can be defined as two sets of itemsets in the form X (Antecedent) → Y (Consequent), and satisfies the condition that $X, Y \subseteq I$ and $X \cap Y = \{\}$.

In the Apriori algorithm, the extraction of association rules is based on parameters, support (S), confidence (C), and lift (L). The support parameter identifies the probability of X, and Y occurring together in the database and is mathematically defined as:

$$\text{Support } (X \rightarrow Y) = P\ (X \cap Y) = \frac{|X \cup Y|}{|D|} \tag{1}$$

Here, $|X \cup Y|$ is the number of times both itemsets X and Y occur together in the database, and |D| represents the total number of items in the specified pedestrian crash database. The confidence parameter measures the probability that an item Y occurs given that an item X occurs, indicating the probability of P (Y|X). In other words, confidence measures the credibility of the association rules and is defined as:

$$\text{Confidence } (X \rightarrow Y) = P\ (Y|X) = \frac{\text{Support } (X \cup Y)}{\text{Support } (X)} = \frac{|X \cup Y|}{|X|} \tag{2}$$

Here | X | is the number of occurrences of only itemset X, and $|X \cup Y|$ is the number of times both itemsets X and Y occur together in the database. Although the support-confidence framework can be used to identify frequent itemsets from the dataset, however, a significant drawback of is that they are satisfied by many association rules, leading to the generation of a lot of uninteresting rules (Weng et al., 2016; Wu et al., 2019). It has been argued that both indicators fail to consider the correlation that exists between numerator and denominator in Equation (2). To address this issue, a more effective indication known as lift has been proposed. Lift considers how much the occurrence probability of Y changes given that X has occurred (Cunjin et al., 2015) and it is defined as:

$$\text{Lift } (X \rightarrow Y) = \frac{\text{Confidence } (X \rightarrow Y)}{\text{Support } (Y)} = \frac{\text{Support } (X \cup Y)}{\text{Support } (X) \times \text{Support } Y} \tag{3}$$

In the above equation, the numerator measures the co-occurrence of itemset X and itemset Y, while the denominator measures the frequency of co-occurrence of itemset X and itemset Y of the rule based on the assumption of conditional independence. When the lift measure is equal to 1, it indicates no correlation between the itemsets X and Y. A lift measure below 1 suggests that the occurrence of itemset X is independent of, or even negatively associated with, the occurrence of itemset Y. Conversely, a lift measure greater than 1 signifies that there is a meaningful association between itemsets X and Y, with higher values indicating a stronger dependency between the two. Since there is no specific rule for setting the minimum threshold of support and confidence parameter, this study gradually changed threshold values until meaningful association rules were found that provide critical insights about crash occurrence.

In addition to the basic support-confidence-lift framework, this study additionally considered Lift Increase Criterion (LIC) measure to identify critical insights, recommended in previous literature (M. M. Hossain et al., 2023; Javed et al., 2026; Montella et al., 2020, 2021). LIC is an incremental measure of how much adding an 'item' to the antecedent improves the predictive power of the rule and shown by the following equation:

$$\text{LIC}(X \rightarrow Y \mid Z) = \frac{\text{Lift}(X \cup Z \rightarrow Y)}{\text{Lift}(X \rightarrow Y)} \quad (4)$$

A value of LIC greater than 1indicates that the inclusion of the additional condition Z enhances the strength of association with Y; a value equal to 1 implies that Z does not alter the association; and a value below 1 suggests that the inclusion of Z weakens the predictive strength of the rule. Note that, the LIC is introduced as an additional exploratory dimension to enrich the association rule analysis. Nevertheless, the primary rule exploration based on support and lift remains fundamental and methodologically sound, as the initial rules identified using these traditional metrics capture the most frequent and reliable crash-risk patterns and thus represent the core structure of the association landscape. The LIC-based assessment of an additional rules is intended to complement this primary layer by highlighting associations that exhibit a meaningful increase in lift, thereby indicating potentially high-impact conditions. Together, these two analytical layers provide a more comprehensive and balanced characterization of crash-risk relationships. The analysis was carried out with the help of the open-source program R version 4.0.1 and the R package 'arules'.

In the context of pedestrian crash analysis, an itemset represents a combination of variables related to a crash event, such as human (driver/pedestrian), vehicle, environmental, roadway, and temporal factors. The association rules generated from these itemsets help identify how certain risk factors co-occur and contribute to crash outcomes. For example, an association rule in this study may show how the absence of crosswalks and poor lighting conditions together increase the likelihood of crashes at midblock locations. The metrics used in ARM, support, confidence, lift and LIC are critical for evaluating the significance and strength of these associations. By utilizing these measures, this study provides a more granular understanding of the association between different crash-contributing factors, allowing for the identification of patterns that traditional statistical models may fail to address. The application of ARM thus fits seamlessly into the broader objective of this research, which is to identify insights for improving pedestrian safety by recognizing the unique combinations of factors that elevate crash risks.

## 4. DATA

### 4.1 Transition in Louisiana Crash Reporting System and Data Inclusion Criteria

The present study utilizes Louisiana's pedestrian crash database from 2017 to 2021, while deliberately excluding data from 2022 onward due to a major change in the state's crash reporting system. Beginning in late 2022, the LaDOTD, in collaboration with the Center for Analytics & Research in Transportation Safety (CARTS), transitioned to a fully electronic crash reporting system known as LA eCrash. This system replaced the previous hybrid paper electronic reporting process and introduced a revised crash report format aligned with the Model Minimum Uniform Crash Criteria (MMUCC) 5th Edition[1]. The statewide rollout of LA eCrash was completed by January 2023[2], making it the official platform for all law enforcement agencies in Louisiana. While this modernization substantially improves data accuracy and reporting efficiency, it also introduces potential discontinuities in variable definitions, coding structures, and data quality relative to pre-2022 records. To maintain methodological consistency and ensure comparability of crash attributes across study years, data from 2022 onward were excluded from the analysis.

---

[1] *Louisiana Highway Safety Improvement Program (HSIP) Annual Report, 2022*

[2] *Louisiana Highway Safety Improvement Program (HSIP) Annual Report, 2023*

The data collection was conducted in two major steps: a) crash-specific data collection, and b) site-specific data collection. Following this, the compilation of crash and site-specific data is also discussed.

**4.2 Crash-specific Data Collection**
Microsoft access crash database provided by the Louisiana Department of Transportation and Development (LaDOTD) was used to compile a list of police-reported pedestrian collisions from 2017 to 2021. During the query design in Microsoft Access software, 'CRASH_NUM' (crash number) was used to merge four data tables including 'Pedestrian table', 'Crash table', 'Vehicle table', and 'DOTD table'. Note that some of the pedestrian crashes had an 'unknown' severity level (total = 285 crashes), which was removed during the data preparation process. The primary database includes 7,399 pedestrian crashes that occurred between 2017 and 2021 in the state of Louisiana. To specify the crash location, a variable 'CRASH_TB_intersection' is available in the crash database with two categories of this variable (TRUE, or FALSE). According to the Louisiana Crash Report Manual (2019), the definition of this variable is provided below (**Figure 2**):

a) at an intersection (CRASH_TB_intersection = TRUE)
b) not at an intersection (CRASH_TB_intersection = FALSE)

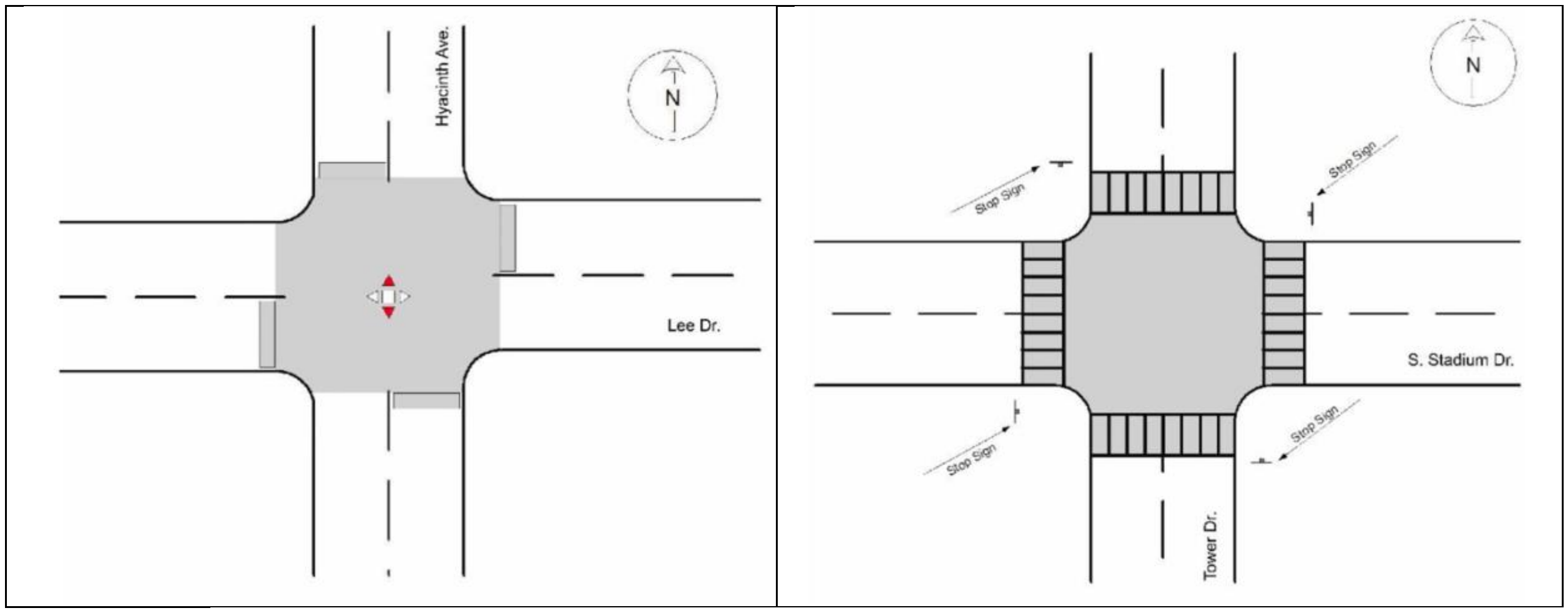

**Figure 2. Intersection crash defined by Louisiana Crash Report Manual (2019)**

Overall, CRASH_TB_intersection = TRUE is designated by the crashes occurring within the physical area of intersections. The other category 'CRASH_TB_intersection = TRUE' suggests that the crash occurred at non-intersection locations and was selected as a filter criterion. At this step, the initial collection of 7,399 pedestrian crashes was further divided into two categories: intersection (3,165, 42.7%) and non-intersection (4,234, 57.3%). The flowchart of the database preparation and analysis framework is shown in **Figure 3**.

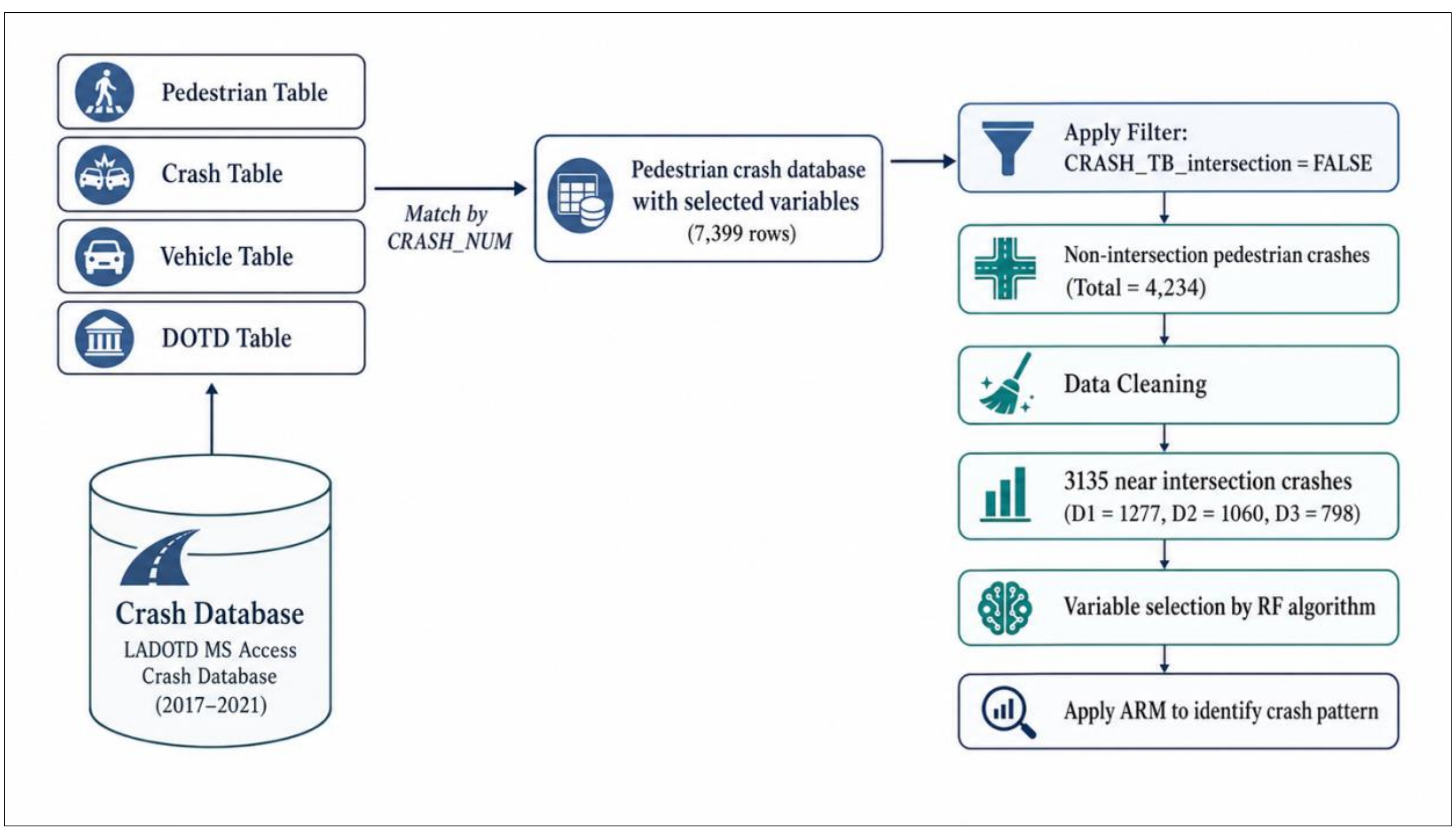

**Figure 3. Data Collection and Analysis Flowchart**

### 4.3 Site-Specific Data Collection

The collection, preparation, cleaning, and segmentation of site-specific crash data was the most critical part of this research. The research team initially started with the 4,234 pedestrian crashes that occurred at non-intersection locations. A total of 56 crash locations did not have any latitude-longitude information and that is why it was initially removed from the database. In the next step, the research team utilized Google satellite maps to go to each of the remaining 4,178 non-intersection crash locations virtually. For each crash location, the distance to the nearest intersection was measured using the 'scale' feature of Google Maps. For example, the following pedestrian crash location (Figure 4) is located around 170 ft. from the nearest intersection.

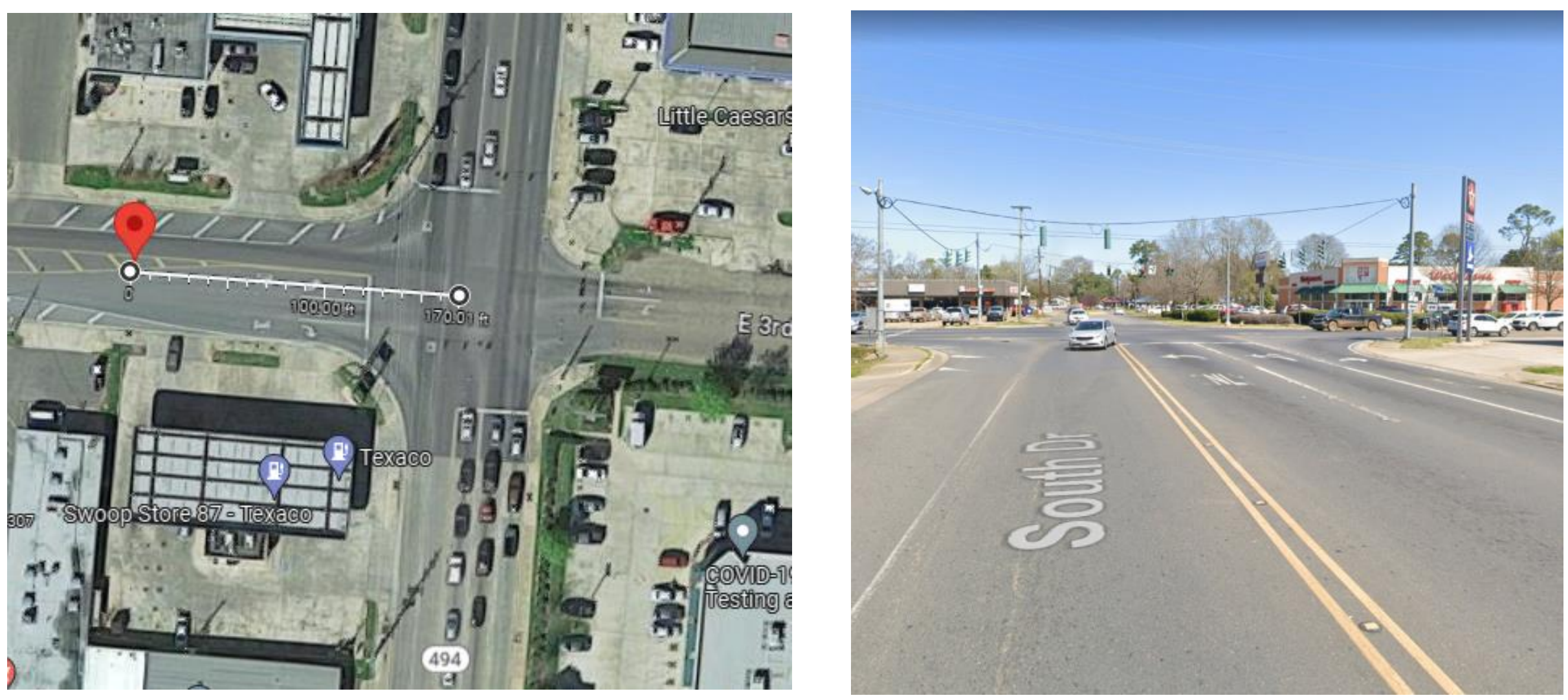

**Figure 4. Example of Site-specific Data Collection**

In a few cases (a total of 637 locations), pedestrian crashes occurred right at the physical area of the intersection (distance to intersection = 0). This could be due to incorrect input by the investigating police officer at the crash scene. These locations were also removed from the crash database. The research team also removed a total of 326 crash locations which were driveways, parking lots, or railway tracks. The primary reason for excluding driveway and parking lot incidents is that the dynamics of vehicle-pedestrian interactions in these locations are fundamentally different from those occurring on roads near intersections. A total of 80 pedestrian crashes occurred on interstate interchanges or interstate highways. Since interstate highways are access-controlled highways, and pedestrians are not usually expected in such locations (e.g., driver's getting out of their disabled vehicles or involved in a previous crash), those crash locations were also removed from the database. The final database consists of 3,135 non-intersection crash locations. Note that, most non-intersection pedestrian crashes in this study occurred on major roadways at 'undesignated' locations, where crosswalks were absent.

### 4.4 Data Segmentation

The research team collected a total of 3,135 pedestrian crashes that occurred at non-intersection locations. It is intuitive that, all these non-intersection pedestrian crashes may not exhibit similar characteristics and that's why a measure of 'distance to intersection' is introduced. The following graph (Figure 5) shows the information on the distance to the nearest intersection and percentile value for these 3,135 pedestrian crashes. Figure 5 can be interpreted as follows: the 50% percentile value was found as 197.6 ft. (~ 198 ft.), which suggests that around 50% of non-intersection pedestrian crashes occurred within 198 ft. of the intersection.

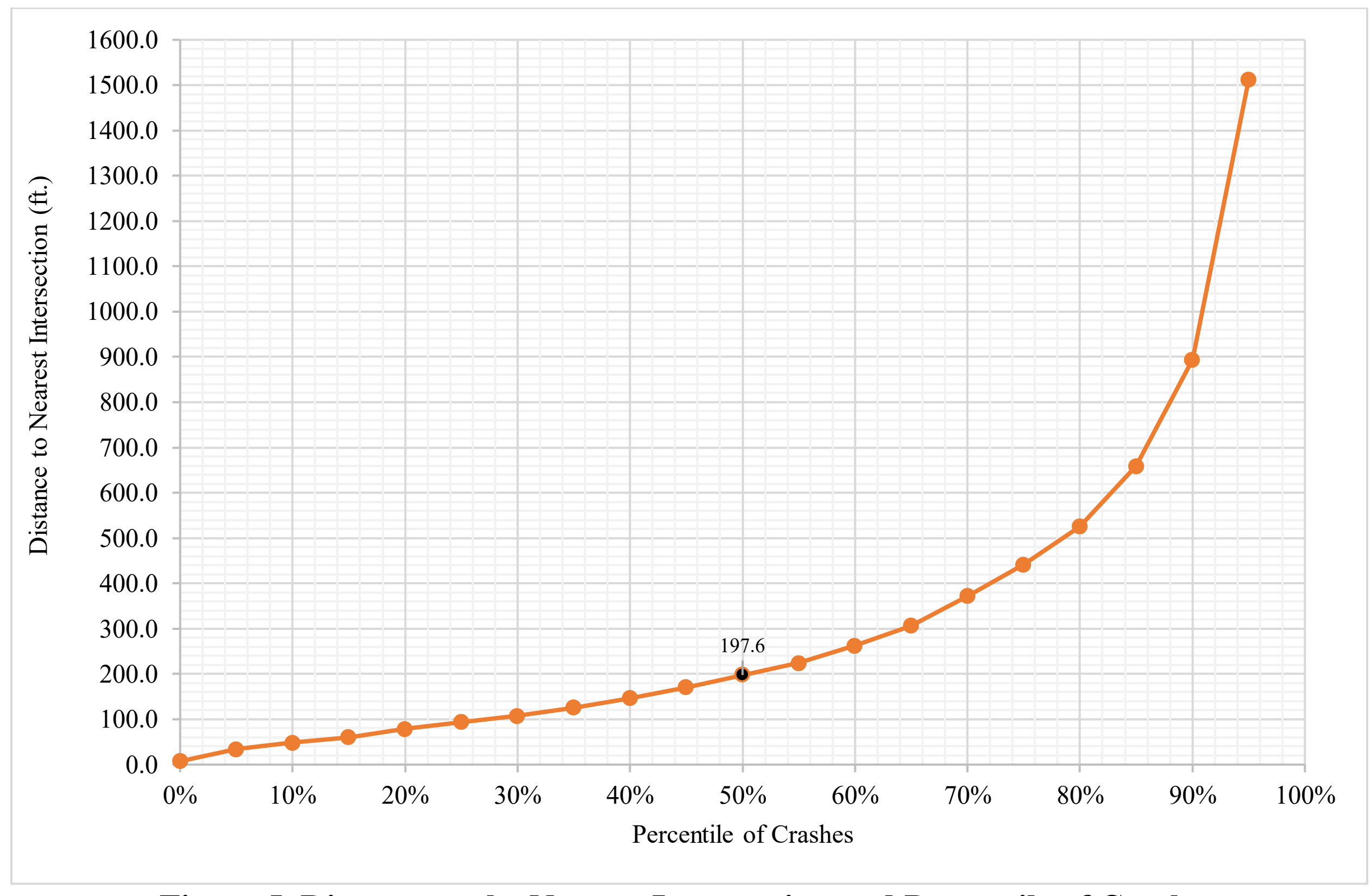

**Figure 5. Distance to the Nearest Intersection and Percentile of Crashes**

For meaningful interpretation and investigation of pedestrian crash patterns, these 3,135 crashes were divided so the data can capture the crash pattern according to distance from the nearest intersection. According to the LaDOTD database, if a crash occurs within a 150 ft. radius of an intersection, then it is designated as an 'intersection-related' crash and falls within the functional area of an intersection. To capture the patterns associated with these crashes, the distance D1 was designed which stands for pedestrian crashes that occurred within 150 ft. of an intersection. According to the database used in this research, around 41% of the pedestrian crashes occurred in zone D1.

A previous research investigated the 'safety influence area' of an intersection and identified approach through volume and posted speed limit as a criterion to address crashes on roadways approaching the intersection (Abdel-Aty & Wang, 2009). This study primarily focused on these two factors as a criterion for our research. In our database, a total of 852 locations had unknown ADT values, and that's why not chosen as a criterion. Focusing on the posted speed of roadways of zone D2 and D3 combined (total = 1,858), the summary of available 1,738 posted speed limit data points is: mean = 38.70, median = 40, and mode = 45. For uncontrolled locations, the driver of approaching vehicles should be able to see pedestrians in adequate time to stop or slow down to avoid a crash. The required sight distance for safe operation at an uncontrolled location is directly related to the vehicle speeds and the distances traveled during perception, reaction, and braking time. According to the American Association of State Highway and Transportation Officials (AASHTO) Green Book (American Association of State Highway and Transportation Officials, 2001), the stopping sight distance is 285 ft. for drivers traveling at 38.7 mph based on the following equation (details in **Table 2)**:

$$SSD = 1.47Vt + \frac{V^2}{30f} \tag{5}$$

Here, V = design speed (mph), t = perception-reaction time (2.5 s, AASHTO standard), f = deceleration factor (0.348, corresponding to AASHTO's deceleration rate of 11.2 ft/s²).

**Table 2. Derivation of Stopping Sight Distance (SSD) at Mean Posted Speed Limit**

| **Component** | **Formula** | **Input Values** | **Result (ft)** |
| --- | --- | --- | --- |
| Perception-reaction distance | $1.47 \times V \times t$ | V = 38.70 mph, t = 2.5 s | 142.2 |
| Braking distance | $V^2 / (30 \times f)$ | V = 38.70 mph, f = 0.348 | 143.5 |
| Total SSD | | | 285.7 ≈ 285 ft |

Keeping this distance in mind, the next threshold distance D2 was designed, and the upper limit was fixed as 435 ft. (150+285). Therefore, the distance D2 designates pedestrian crashes that occur within 151 ft. to 435 ft. of an intersection. The final buffer zone D3 was designed to stand for pedestrian crashes that occurred more than 435 ft. from an intersection. Overall, D1 designates crashes occurring within 150 ft. of an intersection, D2 designates crashes occurring within 151 ft. to 435 ft. of an intersection, and D3 designates crashes occurring at a distance of 435 ft. or higher from an intersection. The collected 3,135 crashes are distributed as follows: D1 (1,277 crashes), D2 (1,060 crashes), and D3 (798 crashes). The framework of buffer zones for distances D1, D2, and D3 is summarized in Figure 6**.**

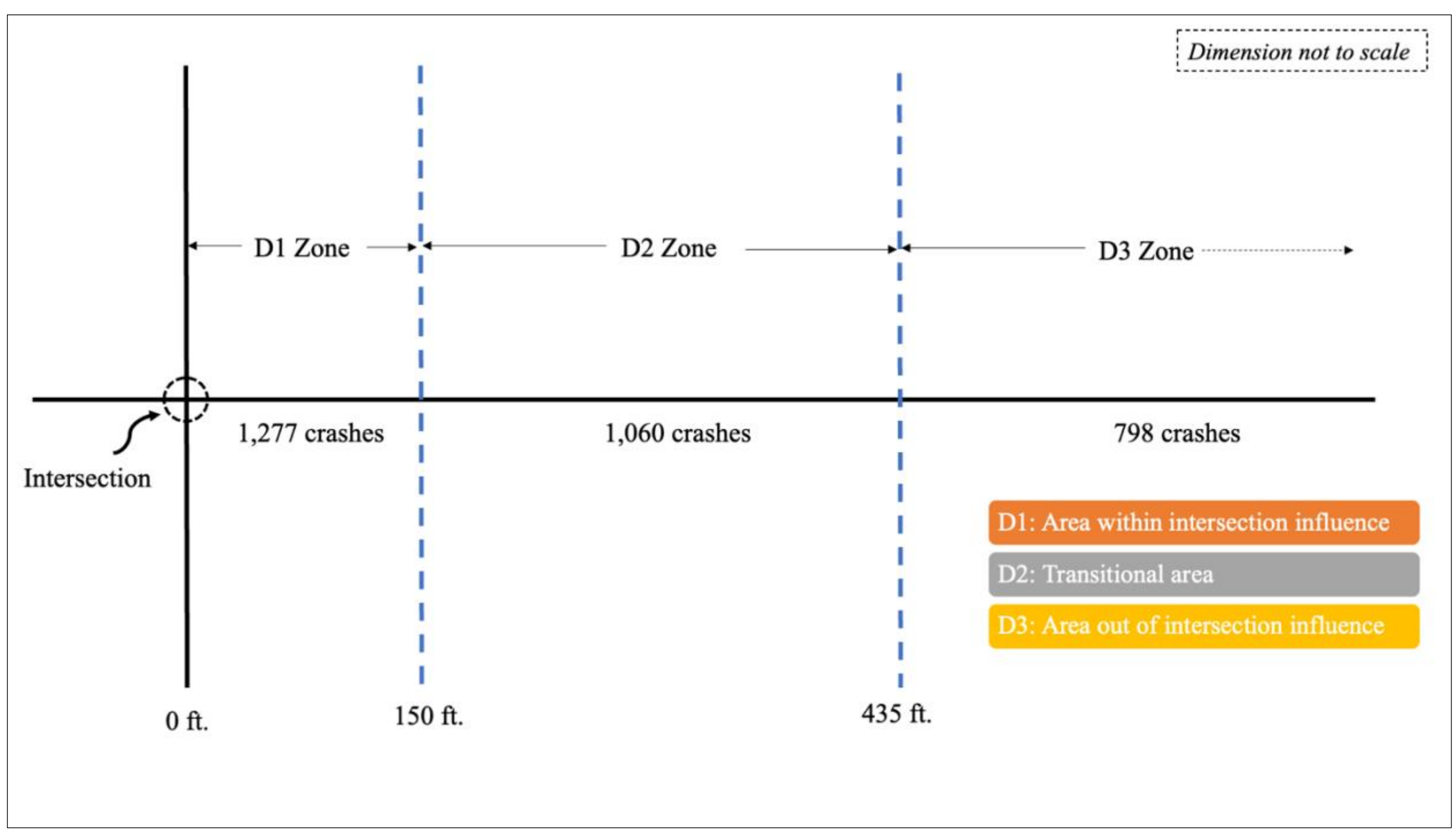

**Figure 6. Framework of Distance to the Intersection**

As shown in the above diagram, the zones D1, D2, and D3 are designed to capture different pedestrian-vehicle crash dynamics based on their distance from the intersection. Zone D1 is characterized as an 'intersection-related' crash zone and is mainly influenced by intersection characteristics and other related factors, such as vehicle maneuvers at the intersection. The decision to keep this zone separate, extending only 150 ft. from the intersection, is rooted in the distinct behavioral patterns of both pedestrians and drivers within proximity to the intersection. This distance is chosen to encapsulate the immediate area where intersection-specific factors dominate crash dynamics, such as turning movements, signal compliance, and crosswalk usage.

Within Zone D1, vehicles are generally not traveling at a steady operating speed. They are accelerating from a stop, decelerating to turn, or yielding to pedestrians and other vehicles. Because the stopping sight distance (SSD) model assumes a vehicle traveling at a constant design speed prior to perceiving a hazard, this assumption does not hold within the intersection-influence area. Therefore, the SSD criterion is not applied within Zone D1, and the 150 ft. distance is instead governed by intersection-specific behavioral factors rather than stopping sight distance.

The SSD criterion becomes relevant once vehicles have cleared the intersection-influence area and resumed a steady travel speed, which this study assumes occurs at the outer edge of Zone D1. Using the mean posted speed limit of 38.7 mph observed across zones D2 and D3, the calculated SSD is 285 ft. This distance is added to the 150 ft. boundary of D1, rather than measured from the intersection itself, because the SSD model applies only to the portion of the roadway where vehicles are traveling at a steady operating speed, which begins at the D1 boundary. This produces a 435 ft. (150 + 285) upper limit for Zone D2, with Zone D3 extending beyond this point. Zone D2 therefore represents a transitional area where both residual intersection influence and stopping sight distance considerations jointly affect crash dynamics, while Zone D3 reflects an area governed primarily by stopping sight distance, with minimal intersection influence.

## 5. RESULTS

This section includes information on the primary variables that were selected, as well as a discussion of the Random Forest algorithm's selection of important variables. With the selected key variables, the results of association rules mining are presented in the following subsection.

### 5.1 Data Exploration

In this study, a total of 21 variables were primarily selected for analysis. Variable name, definition (according to the Louisiana Crash Report Manual), code, and literature reference have been listed in **Table 3**).

**Table 3. Variable information with reference used for analysis**

| **Variable (full name)** | **Definition (Louisiana Crash Report Manual 2019)** | **Code** | **Reference** |
|---|---|---|---|
| Pedestrian gender | Gender of the pedestrian involved in the crash | ped_gender | (Abdullah et al., 2019; Chai et al., 2016) |
| Pedestrian age | Age of the pedestrian involved in the crash | ped_age | (Niebuhr et al., 2016; Prange et al., 2010; Toran Pour et al., 2018) |
| Pedestrian dark clothing | Information about the upper or lower body clothing of the pedestrian (light or dark colored) | ped_dark_cloth | (Black et al., 2023; R. A. Tyrrell et al., 2016) |
| Pedestrian condition | For each pedestrian involved in the crash, enter the letter that best describes his or her condition | ped_condition | (Chaudhari et al., 2021; A. Hossain, Sun, Shahrier, et al., 2023a) |
| Pedestrian alcohol/drug involvement | For all pedestrians involved in the crash, enter the letters that best describe the assessment of whether alcohol or drugs were present in pedestrians and the results of any tests given. | ped_alcohol_drug | (Reish et al., 2021; M. Thomas et al., 2019) |
| Road Type | Enter the letter that best describes the number of lanes, the physical construction, and layout of the roadway at the time and place of the crash. | road_type | (Montella et al., 2011a; Reish et al., 2021) |
| Location Type | Type of location where the crash occurred | location_type | (Pulugurtha & Sambhara, 2011; Ukkusuri et al., 2012) |
| Posted Speed Limit | Posted speed on the road at the crash scene. | posted_speed_limit | (M. Islam, 2023; Nishimoto et al., 2019) |
| Primary Contributing Factor | Causative factors for the crash. | primary_factor | (Moradi et al., 2019; Spainhour et al., 2006) |
| Injury Severity | The injury type most accurately describes the injuries sustained by the pedestrian as a result of this crash. | injury_severity | (Spainhour et al., 2006; Sze & Wong, 2007) |

| **Variable (full name)** | **Definition (Louisiana Crash Report Manual 2019)** | **Code** | **Reference** |
|---|---|---|---|
| Day of week | Month, Day, and Year of the crash (MMDDYYYY). | day_of_week | (Mokhtarimousavi et al., 2020; Song et al., 2021) |
| Crash Hour | Time in hours and minutes when the crash occurred. Write the time using the 24-hour clock where 0000 is midnight and 1200 is noon. | crash_hour | (Gu & Peng, 2021; Sun et al., 2019) |
| Driver Gender | The gender of the driver involved in the crash | driver_gender | (Bener & Crundall, 2008; Montella et al., 2011b) |
| Driver Age | Age of the driver involved in the crash | driver_age | (Kemnitzer et al., 2019; C. Lee & Abdel-Aty, 2005) |
| Driver Condition | For each driver involved in the crash, enter the letter that best describes his or her condition | driver_condition | (Chrysler et al., 2015; Fitzpatrick et al., 2014) |
| Driver Violation | For each driver involved in the crash, enter the letter that best describes a violation by that driver. Choose the factor that most contributed to the crash regardless of whether a citation was issued, or an arrest made as a result of that violation. | driver_violation | (Ahmad & Khattak, 2024; Habibovic et al., 2013) |
| Vehicle Type | Type of vehicle involved in the crash | vehicle_type | (Hu & Cicchino, 2022; Jafari et al., 2025) |
| Presence of Passenger | Number of passengers present in the vehicle while involved in crash (0 if no passenger present) | passenger_presence | (C. Lee & Abdel-Aty, 2008; Orsi et al., 2013) |
| Weather Condition | Enter the letter that best describes the prevailing atmospheric condition that existed at the time and location of the crash. | weather_condition | (Li et al., 2017; Zhai et al., 2019) |
| Lighting Condition | Enter the letter that best describes the lighting conditions that existed at the place and time of the crash. | lighting_condition | (A. Hossain et al., 2022; A. Hossain, Sun, Shahrier, et al., 2023b; A. Hossain, Sun, Thapa, et al., 2023a) |
| Area Type | Area type (urban/rural) in which the pedestrian crash has occurred. This variable is sourced from GIS shape file data. | Rural_Urban | (S. Islam & Jones, 2014b; Sun & Sun, 2020b) |

The variables are summarized according to the measure 'distance_to_intersection' and categorized as D1, D2, and D3 zones (**Table 4**). The first 3 rows (D1 zone) can be interpreted as follows: out of all the 1,277 crashes that occurred in the D1 zone, 65.1% involved male pedestrians, 34.1% involved female pedestrians, and the gender of the pedestrians involved in the remaining 0.9% of the crashes were unknown. Note that the sum may not be exactly 100% column-wise due to rounding errors.

**Table 4. Descriptive Statistics of Key Selected Variables**

| Type | Variable name | Variable Feature | D1 zone (1,277 crashes) | | D2 zone (1,060 crashes) | | D3 zone (798 crashes) | |
|---|---|---|---|---|---|---|---|---|
| | | | # | % | # | % | # | % |
| Pedestrian characteristics | ped_gender | male | 831 | 65.1% | 720 | 67.9% | 578 | 72.4% |
| | | female | 435 | 34.1% | 333 | 31.4% | 216 | 27.1% |
| | | unknown | 11 | 0.9% | 7 | 0.7% | 4 | 0.5% |
| | ped_age | <15y | 189 | 14.8% | 133 | 12.5% | 68 | 8.5% |
| | | 15-24y | 203 | 15.9% | 192 | 18.1% | 138 | 17.3% |
| | | 25-40y | 371 | 29.1% | 312 | 29.4% | 262 | 32.8% |
| | | 41-64y | 389 | 30.5% | 328 | 30.9% | 260 | 32.6% |
| | | >64y | 96 | 7.5% | 71 | 6.7% | 51 | 6.4% |
| | | unknown | 29 | 2.3% | 24 | 2.3% | 19 | 2.4% |
| | ped_dark_cloth | no | 773 | 60.5% | 617 | 58.2% | 462 | 57.9% |
| | | yes | 504 | 39.5% | 443 | 41.8% | 336 | 42.1% |
| | ped_condition | normal | 431 | 33.8% | 352 | 33.2% | 267 | 33.5% |
| | | inattentive_distracted | 371 | 29.1% | 285 | 26.9% | 202 | 25.3% |
| | | alcohol_drug | 98 | 7.7% | 88 | 8.3% | 70 | 8.8% |
| | | illness_fatigued_asleep | 9 | 0.7% | 6 | 0.6% | 2 | 0.3% |
| | | other | 39 | 3.1% | 30 | 2.8% | 28 | 3.5% |
| | | unknown | 329 | 25.8% | 299 | 28.2% | 229 | 28.7% |
| | ped_alcohol_drug | no | 810 | 63.4% | 645 | 60.8% | 482 | 60.4% |
| | | yes | 104 | 8.1% | 96 | 9.1% | 79 | 9.9% |
| | | others | 207 | 16.2% | 185 | 17.5% | 166 | 20.8% |
| | | unknown | 156 | 12.2% | 134 | 12.6% | 71 | 8.9% |
| Road factors | road_type | two_no_separation | 891 | 69.8% | 812 | 76.6% | 625 | 78.3% |
| | | two_physical_separation | 219 | 17.1% | 179 | 16.9% | 149 | 18.7% |
| | | one_way | 131 | 10.3% | 52 | 4.9% | 12 | 1.5% |
| | | other_unknown | 36 | 2.8% | 17 | 1.6% | 12 | 1.5% |
| | location_type | residential | 458 | 35.9% | 445 | 42.0% | 343 | 43.0% |
| | | business_mixed_residential | 423 | 33.1% | 296 | 27.9% | 148 | 18.5% |
| | | business_industrial | 339 | 26.5% | 278 | 26.2% | 178 | 22.3% |
| | | open_country | 29 | 2.3% | 24 | 2.3% | 112 | 14.0% |
| | | other_locality | 28 | 2.2% | 17 | 1.6% | 17 | 2.1% |
| | posted_speed_limit | <30 | 454 | 35.6% | 318 | 30.0% | 145 | 18.2% |
| | | 30-35 | 324 | 25.4% | 250 | 23.6% | 122 | 15.3% |
| | | 40-45 | 284 | 22.2% | 263 | 24.8% | 209 | 26.2% |
| | | 50-55 | 117 | 9.2% | 146 | 13.8% | 255 | 32.0% |
| | | >55 | 3 | 0.2% | 8 | 0.8% | 23 | 2.9% |
| | | unknown | 95 | 7.4% | 75 | 7.1% | 44 | 5.5% |
| Crash factors | primary_factor | ped_violation | 615 | 48.2% | 479 | 45.2% | 325 | 40.7% |
| | | pedestrian_actions | 357 | 28.0% | 297 | 28.0% | 262 | 32.8% |
| | | prior_movement | 175 | 13.7% | 137 | 12.9% | 81 | 10.2% |
| | | driver_condition | 34 | 2.7% | 44 | 4.2% | 35 | 4.4% |
| | | pedestrian_condition | 36 | 2.8% | 31 | 2.9% | 29 | 3.6% |
| | | vision_obscurement | 17 | 1.3% | 15 | 1.4% | 8 | 1.0% |
| | | other_factors | 43 | 3.4% | 57 | 5.4% | 58 | 7.3% |
| | injury_severity | fatal | 121 | 9.5% | 123 | 11.6% | 153 | 19.2% |
| | | severe | 142 | 11.1% | 123 | 11.6% | 86 | 10.8% |
| | | moderate | 498 | 39.0% | 410 | 38.7% | 250 | 31.3% |
| | | complaint | 434 | 34.0% | 338 | 31.9% | 250 | 31.3% |
| | | PDO | 82 | 6.4% | 66 | 6.2% | 59 | 7.4% |
| Temporal Factors | day_of_week | weekday | 929 | 72.7% | 760 | 71.7% | 580 | 72.7% |
| | | weekend | 348 | 27.3% | 300 | 28.3% | 218 | 27.3% |
| | crash_hour | 12am-6am | 113 | 8.8% | 109 | 10.3% | 70 | 8.8% |
| | | 6am-12pm | 209 | 16.4% | 173 | 16.3% | 134 | 16.8% |

| Type | Variable name | Variable Feature | D1 zone (1,277 crashes) | | D2 zone (1,060 crashes) | | D3 zone (798 crashes) | |
|---|---|---|---|---|---|---|---|---|
| | | | # | % | # | % | # | % |
| | | 12pm-6pm | 333 | 26.1% | 246 | 23.2% | 155 | 19.4% |
| | | 6pm-12am | 622 | 48.7% | 532 | 50.2% | 439 | 55.0% |
| Driver factors | driver_gender | male | 620 | 48.6% | 516 | 48.7% | 405 | 50.8% |
| | | female | 425 | 33.3% | 343 | 32.4% | 273 | 34.2% |
| | | unknown | 232 | 18.2% | 201 | 19.0% | 120 | 15.0% |
| | driver_age | 15-24y | 185 | 14.5% | 153 | 14.4% | 149 | 18.7% |
| | | 25-40y | 352 | 27.6% | 275 | 25.9% | 208 | 26.1% |
| | | 41-64y | 360 | 28.2% | 293 | 27.6% | 230 | 28.8% |
| | | >64y | 108 | 8.5% | 113 | 10.7% | 76 | 9.5% |
| | | unknown | 272 | 21.3% | 226 | 21.3% | 135 | 16.9% |
| | driver_condition | normal | 687 | 53.8% | 549 | 51.8% | 427 | 53.5% |
| | | inattentive_distracted | 212 | 16.6% | 175 | 16.5% | 157 | 19.7% |
| | | alcohol_drug | 41 | 3.2% | 38 | 3.6% | 36 | 4.5% |
| | | illness_fatigued_asleep | 10 | 0.8% | 8 | 0.8% | 10 | 1.3% |
| | | other_unknown | 327 | 25.6% | 290 | 27.4% | 168 | 21.1% |
| | driver_violation | no_violations | 674 | 52.8% | 532 | 50.2% | 424 | 53.1% |
| | | careless_operation | 159 | 12.5% | 134 | 12.6% | 100 | 12.5% |
| | | failure_to_yield | 46 | 3.6% | 28 | 2.6% | 21 | 2.6% |
| | | exceeding_speed_limit | 7 | 0.5% | 5 | 0.5% | 7 | 0.9% |
| | | others | 381 | 29.8% | 350 | 33.0% | 241 | 30.2% |
| | | unknown | 10 | 0.8% | 11 | 1.0% | 5 | 0.6% |
| Vehicular factors | vehicle_type | passenger_car | 533 | 41.7% | 448 | 42.3% | 319 | 40.0% |
| | | light_truck | 284 | 22.2% | 248 | 23.4% | 235 | 29.4% |
| | | Van_SUV | 336 | 26.3% | 254 | 24.0% | 170 | 21.3% |
| | | other | 75 | 5.9% | 62 | 5.8% | 46 | 5.8% |
| | | unknown | 49 | 3.8% | 48 | 4.5% | 28 | 3.5% |
| | passenger_presence | no | 1069 | 83.7% | 895 | 84.4% | 658 | 82.5% |
| | | yes | 208 | 16.3% | 165 | 15.6% | 140 | 17.5% |
| Environmental factors | weather_condition | clear | 1001 | 78.4% | 817 | 77.1% | 623 | 78.1% |
| | | cloudy | 189 | 14.8% | 161 | 15.2% | 112 | 14.0% |
| | | rain | 69 | 5.4% | 64 | 6.0% | 42 | 5.3% |
| | | fog_sleet_snow | 8 | 0.6% | 10 | 0.9% | 11 | 1.4% |
| | | other_unknown | 10 | 0.8% | 8 | 0.8% | 10 | 1.3% |
| | lighting_condition | daylight | 589 | 46.1% | 458 | 43.2% | 294 | 36.8% |
| | | dark_with_streetlight | 463 | 36.3% | 346 | 32.6% | 155 | 19.4% |
| | | dark_no_streetlight | 176 | 13.8% | 222 | 20.9% | 311 | 39.0% |
| | | dusk_dawn | 40 | 3.1% | 28 | 2.6% | 26 | 3.3% |
| | | other_unknown | 9 | 0.7% | 6 | 0.60% | 12 | 1.5% |
| | Rural_Urban | rural | 299 | 23.4% | 280 | 26.4% | 394 | 49.4% |
| | | urban | 978 | 76.6% | 780 | 73.6% | 404 | 50.6% |

### 5.2 Variable Selection by Random Forest Algorithm

An appropriate choice of variables is one of the most important components in using the ARM algorithm to produce logical and meaningful association rules. The incorporation of noisy variables in the model could produce extremely complex 'decision rules' that are challenging to comprehend (Ait-Mlouk, Agouti, et al., 2017). The study utilized the Random Forest (RF) algorithm as a method of variable selection (setting 'distance_to_intersection' as the response variable and all other variables as explanatory). Mean Decrease Gini (MDG) was used as a measure for assessing feature importance. Figure 7 shows the variable importance plot where each variable is shown on the y-axis and associated MDG on the x-axis. It is worth noting that variables on the y-axis are ordered from most to least important (top to bottom). The top 14 variables accounted for 85% of cumulative MDG values, and the variables were selected based on this cumulative threshold of MDG. Note that there is no specific standard for variable selection according to MDG values. A similar threshold has been adopted by a few of the previous studies (A. Hossain et al., 2022; Sun et al., 2022).

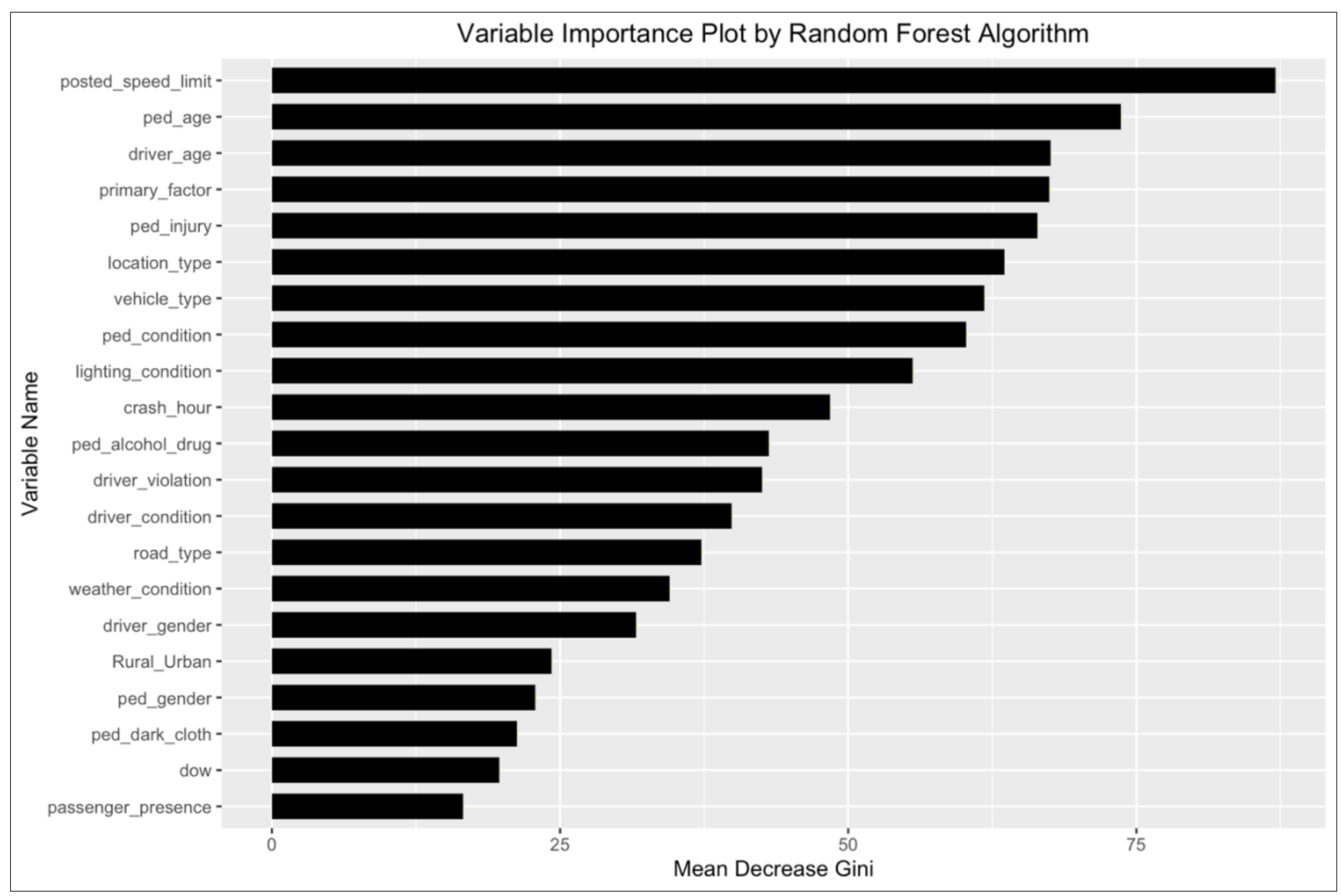

**Figure 7. Variable selection by Random Forest Algorithm**

### 5.3 Results of Association Rules Mining

Under this subsection, the setup for the ARM algorithm is first discussed, and the results are presented according to D1, D2, and D3 zones, respectively.

#### *5.3.1 ARM Model Framework*

ARM algorithms typically require transaction data in a specific format known as a transaction database (3,135×73), where 3,135 is the number of unique pedestrian crashes, and 73 is the total number of variable features. In order to achieve the desired outcome when taking into account a large-sized matrix (235,125 cell entries in this case), the identification of meaningful association rules necessitates the intuitive handling of model parameters (support, confidence, and lift) and setting their minimum threshold values (trial-and-error process). To capture meaningful crash patterns, association rules were selected based on a specified minimum threshold of support (S) and confidence (C) values. A similar trial-and-error process has been extensively used in previous research using the ARM algorithm (Ait-Mlouk, Agouti, et al., 2017; Das et al., 2019a, 2019b; Das, Kong, et al., 2021; Das, Tamakloe, et al., 2021; Rahman et al., 2021).

*5.3.1.1 Rules Generation by Lift and Support*

To capture association rules with high-frequency and high-confidence occurrence, every 20 sets of association rules cover high-support and confidence rules (**Table 5**, **Table 7**, **Table 9**). For example, R1-R10 are chosen based on lift value, while R11-R20 are chosen based on support value. The overall idea is to capture the trade-off in selecting association rules from two different contexts: a) low support but high confidence, and b) high support but low confidence. High-confidence, low-support rules, often emerge from specific and rare conditions within the dataset. While these rules may indeed be based on a smaller number of observations, their high confidence indicates a strong, reliable relationship between the antecedent and the consequent when the conditions do occur. This makes them particularly valuable for identifying high-risk scenarios that, although rare, are almost always more likely to occur in specific zones. On the other hand, high-support, low-confidence rules reflect broader trends within the dataset. These rules are applicable to a larger portion of the data, capturing more common conditions that might lead to crashes in specific zones. Both types of rules are essential for a comprehensive understanding of crash dynamics and for developing a balanced approach to traffic safety interventions.

Note that the minimum threshold for lift value is considered as 1.1 to generate stronger rules for all cases. For ease of interpretation, the rules generation was limited to 4-itemset rules. Using the selected variables, ARM was applied to three separate scenarios: distance_to_intersection = D1, distance_to_intersection = D2, and distance_to_intersection = D3.

*5.3.1.2 Rules Generation by LIC*

Focusing on the LIC measure, rules were generated for 2-itemset, 3-itemset, 4-itemset and 5-itemset based on specified threshold of support and confidence parameter for each individual scenario (e.g., D1, D2, and D3). To generate strong association rules, the confidence parameter was kept at 50% threshold, and support values for each itemset rule generation were set by a trial-and-error process with varying levels (e.g., 0.01, 0.001, 0.0001, 0.00001). The overall idea was to generate around at-most 10,000 rules incorporating efficient model runs. Handling rules generation, and identification of interesting and intuitive rules based on the LIC measure is a critical decision-making task. A few additional conditions were imposed to identify the rules based on LIC measure. To be considered as a mother rule, the count filter was provided priority to generate strong rules. Based on the mother rule, the child rules were also selected based on the count filter as a priority. Child rules with any item reported as 'other/unknown' were not selected to improve interpretation. Finally, the minimum threshold for LIC value was kept 1.05 (minimum threshold) based on recommendations from previous literature.

### 5.4 Crashes Associated with D1 zone (intersection-related)

The 'distance_to_intersection' variable was set to 'D1' as the 'consequent' or Left-Hand-Side (L.H.S) to mine the association rules for case 1. The minimum support and confidence values were set at 0.3% and 50% respectively. Initially, the algorithm generated 3,040 rules which contained many redundant rules. After pruning redundant rules, 1,131 rules remained. **Table 5** lists the top 20 rules (R1- R20) for this case. In addition, a graphical representation of rules R1-R20 is provided in **Figure 8**.

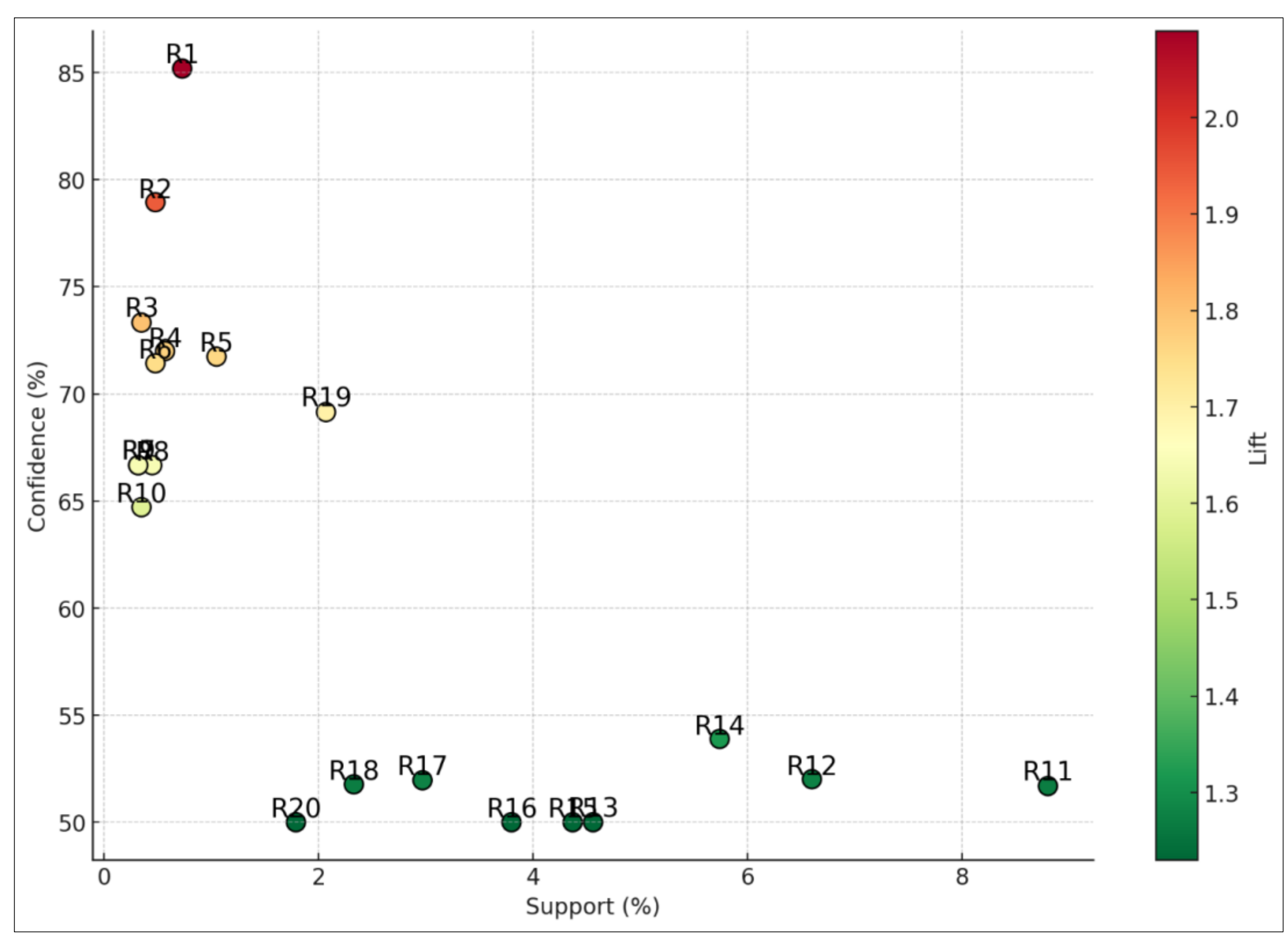

**Figure 8. Top 20 Association Rules for D1 Zone**

**Table 5. Pedestrian crash pattern at D1 zone (within 150 ft. of an intersection)**

| **Selected top rules chosen based on lift value (R1-R10)** | | | | | |
|---|---|---|---|---|---|
| ID | Antecedent (s) or L.H.S of association rules | S (%) | C (%) | L | # |
| R1 | {road_type=one_way, ped_condition=inattentive_distracted} | 0.73 | 85.19 | 2.09 | 23 |
| R2 | {ped_injury=severe, lighting_condition=dark_with_streetlight, vehicle_type=light_truck} | 0.48 | 78.95 | 1.94 | 15 |
| R3 | {ped_injury=severe, crash_hour=12pm-6pm, driver_age=41-64y} | 0.35 | 73.33 | 1.80 | 11 |
| R4 | {ped_injury=fatal, primary_factor=ped_violation, vehicle_type=Van_SUV} | 0.57 | 72.00 | 1.77 | 18 |
| R5 | {primary_factor=pedestrian_actions, road_type=one_way} | 1.05 | 71.74 | 1.76 | 33 |
| R6 | {location_type=business_mixed_residential, driver_violation=failure_to_yield} | 0.48 | 71.43 | 1.75 | 15 |
| R7 | {ped_injury=complaint, location_type=business_industrial, driver_violation=failure_to_yield} | 0.32 | 66.67 | 1.64 | 10 |
| R8 | {lighting_condition=dark_with_streetlight, ped_condition=alcohol_drug, vehicle_type=light_truck} | 0.45 | 66.67 | 1.64 | 14 |
| R9 | {ped_injury=moderate, ped_age=15-24y, driver_age=>64y} | 0.32 | 66.67 | 1.64 | 10 |
| R10 | {ped_age=15-24y, ped_condition=inattentive_distracted, driver_age=>64y} | 0.35 | 64.71 | 1.59 | 11 |
| **Selected top rules based on support value (R11-R20)** | | | | | |
| R11 | {lighting_condition=daylight, posted_speed_limit=<30} | 8.80 | 51.69 | 1.27 | 276 |
| R12 | {primary_factor=ped_violation, location_type=business_mixed_residential} | 6.60 | 52.01 | 1.28 | 207 |
| R13 | {ped_injury=moderate, crash_hour=6pm-12am, lighting_condition=dark_with_streetlight} | 4.56 | 50.00 | 1.23 | 143 |
| R14 | {lighting_condition=dark_with_streetlight, location_type=business_mixed_residential} | 5.74 | 53.89 | 1.32 | 180 |
| R15 | {crash_hour=6pm-12am, lighting_condition=dark_with_streetlight, ped_age=41-64y} | 4.37 | 50.00 | 1.23 | 137 |
| R16 | {crash_hour=6pm-12am, primary_factor=pedestrian_actions, lighting_condition=dark_with_streetlight} | 3.80 | 50.00 | 1.23 | 119 |
| R17 | {ped_injury=moderate, ped_age=<15y} | 2.97 | 51.96 | 1.28 | 93 |
| R18 | {primary_factor=ped_violation, ped_age=<15y} | 2.33 | 51.77 | 1.27 | 73 |
| R19 | {primary_factor=ped_violation, road_type=one_way} | 2.07 | 69.15 | 1.70 | 65 |
| R20 | {ped_age=<15y,ped_condition=inattentive_distracted, vehicle_type=passenger_car} | 1.79 | 50.00 | 1.23 | 56 |

**Note: S = Support, C = Confidence, L = lift, L.H.S = Left-Hand-Side of association rules, # indicates number of occurrences in the database*

Some of the major crash patterns are discussed under the following subheadings:

*5.4.1 Pedestrian Characteristics*

The first rule (R1, L = 2.09) suggests that 'inattentive/distracted' pedestrians are more likely (2.09 times) to be involved in crashes on a one-way road in D1 zones. In general, one-way streets are expected to simplify the crossing behavior for pedestrians, as they must look for traffic in one direction only. In such settings, there is also a safety concern for pedestrians as there is an increased confidence level (leading to inattentiveness or distraction) on the part of pedestrians and higher movement speeds on the part of the driver (David, 1994). When pedestrians are inattentive or distracted, their level of awareness and attention to their surroundings decreases. They may not notice important cues, such as traffic signals, oncoming vehicles, or potential hazards, which can lead to crashes. Additionally, pedestrians might attempt to cross the one-way road in an unexpected direction, assuming that traffic is coming from only one direction. This wrong-way crossing can surprise drivers, increasing the chances of a collision. The other evidence of 'inattentive/distracted' pedestrians was also identified in the D1 zone – [R10] 15-24 years old pedestrians involved in crashes and the driver age was >64 years; [R20] less than 15 years old pedestrians involved in crashes with a passenger car. Overall, young pedestrians (<15 years, 15-24 years) were more likely to be involved in crashes at a closer distance (<150 ft.) to the intersection. With the increasing prevalence of smartphones, tablets, and other electronic devices, young pedestrians may be more likely to be distracted while walking and get involved in crashes in the D1 zone.

The variable attribute 'pedestrian violation' was identified in several instances – [R4] involving Van/SUV type vehicles and resulting in pedestrian fatalities (L = 1.77); [R12] business with mixed residential areas (L = 1.28); [R18] involving pedestrians of age group less than 15 years and [R19] crashes occurring on one-way roads (L = 1.70). Rule R12 states that 'pedestrian violation' was cited as a factor contributing to crashes in the D1 zone in business with mixed residential areas. The mixed-use areas often attract higher pedestrian volumes, especially during peak hours. With more people walking around, there's a higher likelihood of pedestrian violations occurring simply due to the larger numbers. In some mixed-use areas, the infrastructure might not be optimized for pedestrian safety. The absence of designated crosswalks can result in jaywalking or other violations as pedestrians try to navigate busy streets at a closer distance to the intersection. Two of the association rules (R5, R16) have identified 'pedestrian action' as one of the factors contributing to crashes in the D1 zone. Rule R16 (S = 3.8%, L = 1.23) reveals a critical link between temporal (6 pm – 12 am), environmental (dark-with-streetlight), and human factors (pedestrian action).

*5.4.2 Driver characteristics*

Two of the associations identified drivers' 'failing to yield' as a factor contributing to pedestrian crashes at the D1 zone, with both rules potentially suggesting the impact of settings (R6 = business with mixed residential areas; R7 = business/industrial areas) in such crashes. From the driver's perspective, they may not be fully aware of the rules and regulations regarding pedestrian right-of-way at intersections. A typical example can be a driver taking a right turn at the intersection while the pedestrian walk signal is on. Drivers may also misjudge the speed at which pedestrians are crossing the intersection and 'failure to yield' to pedestrians. Two of the other association rules [R9, R10] reveal the involvement of older drivers (>64 years) in pedestrian crashes in the D1 zone. The critical link between 'older drivers' and 'intersection crashes' has been well documented in previous research (Alshehri & Dissanayake, 2022; Braitman et al., 2007; Broberg et al., 2008; Dotzauer et al., 2013; Dukic & Broberg, 2012; Hallmark & Mueller, 2004). Older drivers may experience a decline in peripheral vision, making it difficult for them to spot pedestrians or vehicles approaching from the sides, especially in complex situations near intersections.

*5.4.3 Lighting condition*

Lighting conditions can play a critical role in pedestrian crashes occurring in the D1 zone. Note that streetlights are typically installed at intersections and therefore location D1 is expected to have similar characteristics. Although with the presence of streetlight at the D1 zone, there were several instances in which pedestrians were involved in crashes at dark-with-streetlight conditions – R2 (involving light truck and resulting in severe injury of pedestrians), R8 (involving impaired pedestrians and light truck vehicle type), R13 (during 6 pm – 12 am and resulting in moderate injury crashes), R14 (crashes occurring in business with mixed residential areas), R15 (during 6 pm – 12 am and involving 41-64 years age group pedestrians) and R16 (during 6 pm – 12 am and occurred due to pedestrian actions). Streetlights can create a false sense of security for pedestrians, making them believe they are more visible to drivers than they are. As a result, some pedestrians might be less cautious and more likely to engage in unsafe behavior, such as jaywalking or crossing without checking for oncoming traffic at a distance closer to the intersection.

*5.4.4 Posted Speed Limit*

The rule R11 (lighting_condition = daylight, posted_speed_limit =<30) has a support value of 8.8%, suggesting that there were 276 instances (8.8% multiplied by 3,135) in which pedestrians were involved in crashes during daylight and the posted speed of the approaching roadway to the intersection was 30 mph or lower. The D1 zone likely corresponds to areas with higher pedestrian activity, such as residential neighborhoods, commercial districts, or school zones. These areas are often designed with lower speed limits to enhance pedestrian safety, but the increased interaction between pedestrians and vehicles may inadvertently raise the likelihood of conflicts, especially during daylight hours when pedestrian activity peaks. Note that, this critical combination is 1.27 times more likely to occur in the D1 zone. This may be due to the increased activity and exposure of pedestrians in low-speed settings (e.g., residential, and commercial areas) (Hanson et al., 2013). A lower posted speed limit does not guarantee that drivers adhere to the speed limit (Davis, 2001; McMahon et al., 1999). Furthermore, lower speed limits are often found in areas with complex traffic environments, where drivers may need to navigate crosswalks, intersections, and other potential points of conflict with pedestrians. While lower speeds are intended to reduce the severity of crashes, they do not necessarily eliminate the possibility of crashes occurring, particularly in zones with high pedestrian density. Additionally, daylight conditions might lead to increased pedestrian activity, as more people are out walking during the day, increasing exposure and the chance of an incident. In low-speed zones, pedestrians might assume that drivers will always obey the lower speed limits and be more cautious. This false sense of security can lead pedestrians to take more risks and be less attentive when crossing the road during daylight. Drivers and pedestrians in low-speed zones might become complacent, thinking that the lower speeds make crashes less likely. As a result, both parties might be less vigilant and more prone to making mistakes, increasing the likelihood of crashes.

### 5.5 Pattern Identification by LIC Measure for D1 zone

**Table 6** provides details of rules generation for zone D1. Association rules were generated for 2 to 5-itemsets using several minimum support levels while fixing confidence at 0.5, so only rules with at least 50% reliability were retained. The support parameter was progressively lowered only until the number of rules either approached about 10,000 or stopped increasing further. The 2-itemsets yielded just 2–3 rules at all tested support parameter, indicating that lowering support value did not uncover additional strong pairwise relationships. For 3-itemsets, reducing support value increased the rules up to 452, beyond which further lowering support value provided no gain in rule count. In contrast, 4 and 5-itemsets showed rapid growth in rules as support value decreased, so their minimum support values were chosen at the point where rule counts rose to around 10,000 balancing pattern discovery with computational tractability in the mining process.

**Table 6. Details of Rules Generation for Each Itemset Category (D1)**

| Itemset | Support (S) | Total rules generation with C = 0.5 |
|---|---|---|
| 2-itemset | 0.01 | 2 |
| | 0.001 | 3 |
| | 0.0001 | 3 |
| 3-itemset | 0.01 | 98 |
| | 0.001 | 316 |
| | 0.0001 | 452 |
| | 0.00001 | 452 |
| 4-itemset | 0.01 | 785 |
| | 0.001 | 5689 |
| | 0.0001 | 12601 |
| | 0.0032 | 8785 |
| 5-itemset | 0.01 | 1517 |
| | 0.001 | 35,714 |
| | 0.0035 | 10,514 |

Based on an extensive review process, 52 rules (L1 – L52) were retained for subsequent analysis (details are provided in **Table 11** in Appendix section). Association rules were categorized under the LIC class only if they met a minimum threshold of 1.05, indicating that the occurrence of the outcome was at least 5% more likely when the antecedent conditions were present than when they were not. The following diagram shows details of the 52 rules for zone D1.

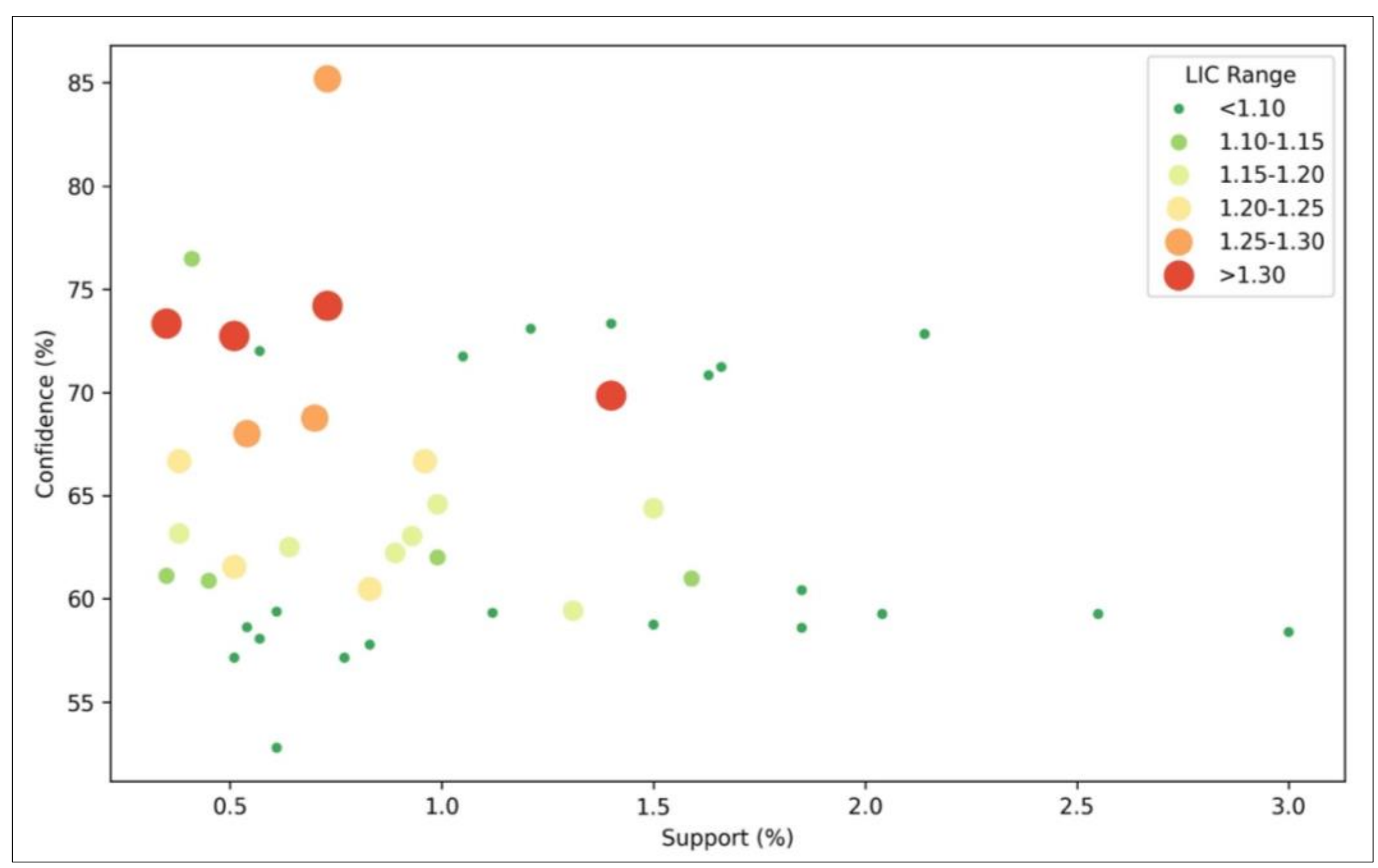

**Figure 9. Details of Association Rules (L1-L52) for zone D1**

The five mother rules establish distinct foundational patterns for pedestrian crashes within 150 feet of intersections (D1 zone). L1 (one-way roads, L = 1.65) represents the strongest single-factor predictor, L11 (low-speed zones with vans/SUVs, L = 1.38) captures vehicle-infrastructure interaction, L23 (dark conditions with streetlights in business/mixed residential areas, L = 1.32) reflects environmental complexity, L37 (child pedestrians with non-violating drivers, L = 1.23) highlights vulnerable user patterns, and L42 (L23's conditions with two-way undivided roads, L = 1.34) demonstrates how road configuration modifies environmental risk.

Mother rule L1's originated rules (L2 – L10) reveal how one-way streets at intersections interact with behavioral and demographic factors. For example, pedestrian inattention (L2, L = 2.09, LIC = 1.27) as a factor, indicating that one-way configurations create false confidence among distracted pedestrians despite critical decision points within 150-foot intersection approach. PDO crashes (L3), middle-aged pedestrians (L4), afternoon periods (L5), low-speed limits (L6), late-night hours (L7), pedestrian-caused factors (L8), and varying injury severities (L9-L10) all show moderate lift increases (LIC = 1.05-1.14), demonstrating that one-way geometry creates consistent D1 patterns across temporal and causal dimensions - likely because simplified directional flow encourages higher approach speeds toward intersections where pedestrian crossings concentrate.

Mother rule L11's originated rules (L12-L22) show that vans/SUVs in low-speed zones compound D1 risk through specific combinations, with one-way configuration (L12, LIC = 1.32) and physical separation (L13, LIC = 1.30) producing the strongest enhancements. This indicates larger vehicles with restricted sightlines become particularly hazardous when road design either simplifies vehicle flow or creates barriers obscuring pedestrians at crossing points. Business/mixed residential locations (L14), afternoon periods (L15), vulnerable populations (children L16, elderly L17), and driver characteristics (L19-L22) show LIC values of 1.05-1.15, demonstrating that SUV/van geometry systematically concentrates crashes near intersections across diverse scenarios – consistent with research showing these vehicles create detection failures precisely where pedestrians make crossing decisions.

Mother rules L23, L37, and L42 reveal environmental and behavioral complexities. L23's originated rules (L24-L36) show dark streetlight conditions in business areas become most hazardous with careless driving (L24, LIC=1.36), pedestrian inattention (L26, LIC = 1.30), and one-way streets (L27, LIC = 1.28). This potentially indicates that even dark-with-streetlight may create challenging visibility in the 150-foot zone where both users must judge distances at night. L37's originated rules (L38-L41) demonstrate that child pedestrians with non-violating drivers show elevated D1 risk during morning hours (L38, LIC = 1.23), dark conditions (L39, LIC = 1.21), and business areas (L40, LIC = 1.19) – reflecting school trips and children's limitations in judging vehicle speeds near intersections. L42's originated rules (L43-L52) reveal that dark conditions on two-way undivided roads in business areas are more likely to create risky situations in D1 zones when combined with young drivers (L43, LIC = 1.33), prior movement factors (L44, LIC = 1.25), and impairment or inattention (L45-L48, LIC = 1.16-1.22). This shows how absence of physical separation increases cognitive demands on drivers approaching intersections at night, with failures manifesting as crashes within 150 ft. of the intersection.

### 5.6 Crashes Associated with D2 zone

To mine the association rules for case 2, the 'distance_to_intersection' variable was set to 'D2' as a 'consequent' in the model. After multiple rounds of trial and error, the minimal support value (0.4%) and confidence (40%) were selected. The algorithm initially generated 1,880 rules, and after pruning, 968 rules remained. **Table 7** lists the top 20 rules (R21-R40) for this case. In addition, the rules are graphically represented in **Figure 10**.

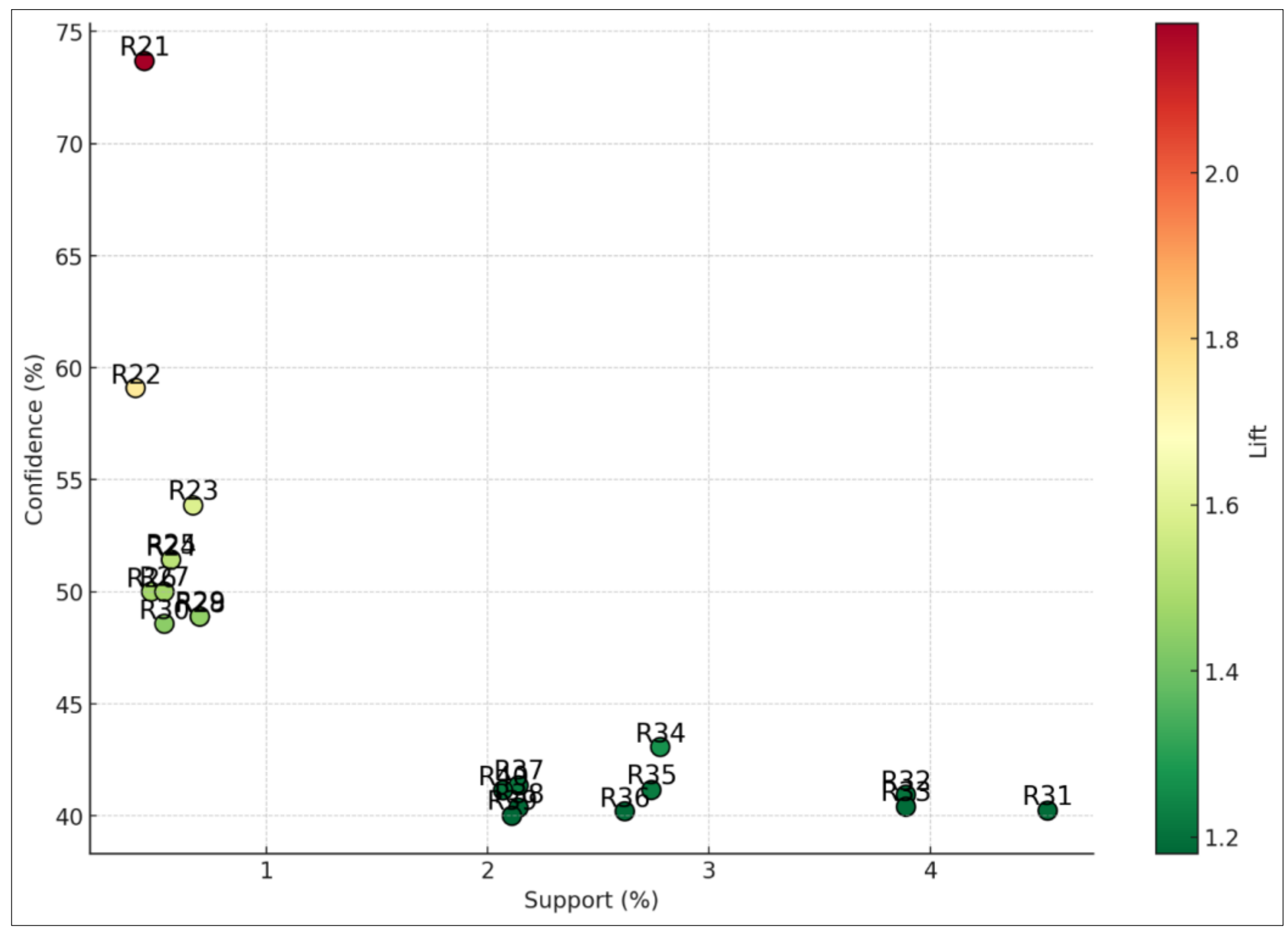

**Figure 10. Top 20 Association Rules for D2 Zone**

**Table 7. Pedestrian Crash Pattern at D2 Zone (within 151 - 435 ft. of an intersection)**

| **Selected top rules based on Lift value (R21-R30)** | | | | | |
|---|---|---|---|---|---|
| ID | Antecedent | S (%) | C (%) | L | # |
| R21 | {primary_factor = driver_condition, driver_age = >64y} | 0.45 | 73.68 | 2.18 | 14 |
| R22 | {road_type = two_physical_separation, posted_speed_limit = <30, vehicle_type = passenger_car} | 0.41 | 59.09 | 1.75 | 13 |
| R23 | {crash_hour = 12am-6am, primary_factor = prior_movement} | 0.67 | 53.85 | 1.59 | 21 |
| R24 | {ped_injury=severe, driver_violation=careless_operation} | 0.57 | 51.43 | 1.52 | 18 |
| R25 | {ped_injury=severe, lighting_condition=dark_no_streetlight, posted_speed_limit=40-45} | 0.57 | 51.43 | 1.52 | 18 |
| R26 | {ped_age=25-40y, ped_condition=alcohol_drug, driver_age=41-64y} | 0.48 | 50.00 | 1.48 | 15 |
| R27 | {ped_injury=fatal, location_type=business_mixed_residential, vehicle_type=light_truck} | 0.54 | 50.00 | 1.48 | 17 |
| R28 | {ped_alcohol_drug=yes, ped_age=41-64y, posted_speed_limit=40-45} | 0.70 | 48.89 | 1.45 | 22 |
| R29 | {lighting_condition=dark_no_streetlight, location_type=business_industrial, ped_age=41-64y} | 0.70 | 48.89 | 1.45 | 22 |
| R30 | {ped_injury=fatal, posted_speed_limit=40-45, vehicle_type=Van_SUV} | 0.54 | 48.57 | 1.44 | 17 |
| **Selected top rules based on support value (R31-R40)** | | | | | |
| R31 | {road_type = two_no_separation, posted_speed_limit=<30, vehicle_type = passenger_car} | 4.53 | 40.23 | 1.19 | 142 |
| R32 | {lighting_condition = dark_with_streetlight, location_type = residential} | 3.89 | 40.94 | 1.21 | 122 |
| R33 | {crash_hour=6pm-12am, location_type=residential, vehicle_type=passenger_car} | 3.89 | 40.40 | 1.19 | 122 |
| R34 | {primary_factor=ped_violation, location_type=business_industrial, road_type=two_no_separation} | 2.78 | 43.07 | 1.27 | 87 |
| R35 | {driver_age=41-64y, posted_speed_limit=40-45} | 2.74 | 41.15 | 1.22 | 86 |
| R36 | {lighting_condition=daylight, location_type=business_industrial, road_type=two_no_separation} | 2.62 | 40.20 | 1.19 | 82 |
| R37 | {crash_hour=12am-6am, lighting_condition=dark_with_streetlight} | 2.14 | 41.36 | 1.22 | 67 |
| R38 | {primary_factor=prior_movement, location_type=residential} | 2.14 | 40.36 | 1.19 | 67 |
| R39 | {location_type=residential, road_type=two_no_separation, posted_speed_limit=30-35} | 2.11 | 40.00 | 1.18 | 66 |
| R40 | {location_type=business_mixed_residential, road_type=two_no_separation, vehicle_type=Van_SUV} | 2.07 | 41.14 | 1.22 | 65 |

**Note: S = Support, C = Confidence, L = lift, L.H.S = Left-Hand-Side of association rules, # indicates number of occurrences in the database*

Some of the major crash patterns are discussed under the following subheadings:

*5.6.1 High-speed roadways (40-45 mph)*

There was evidence of pedestrian crashes at D2 zones that were characterized by high posted speed limits (40-45 mph). Some of the identified rules were – [R25] crashes occurring in dark-no-streetlight conditions resulting in pedestrian severe injury (L = 1.52); [R28] involving alcohol-impaired pedestrians of age group 41-64 years (L = 1.45), [R30] involving Van/SUV type vehicle and resulting in fatal injury of pedestrians (L = 1.44); [R35] involving drivers of age group 41-64 years (L = 1.22). Note that pedestrians are typically not expected by drivers in such locations (D2 zone). High-speed roadways typically have fast-moving vehicles, making it more challenging for drivers to spot and react to pedestrians in time, especially if they suddenly enter the road. Additionally, high-speed roadways may have limited designated crossing points or pedestrian infrastructure, leading pedestrians to attempt unsafe crossings or jaywalking.

*5.6.2 Two-way with no physical separation*

Roadway type may affect pedestrian crashes occurring in D2 zone and were characterized by 'two-way with no physical separation'. Four of the association rules has identified this issue with a complex interaction with 'location type' – [R31] involving passenger car and locations with posted speed limit of 30 mph or lower (L = 1.19), [R34] crashes occurring due to pedestrian violation in business/industrial areas (L = 1.27), [R36] crashes occurring in the daylight condition in business/industrial areas (L = 1.19), [R39] crashes occurring in residential areas with a posted speed limit of 30-35 mph (L = 1.18) and [R40] crashes occurring in business with mixed residential areas and involving Van/SUV type vehicle (L = 1.22). Without a physical barrier between traffic lanes, pedestrians may have difficulty accurately judging the speed and distance of oncoming vehicles (both directions), leading to misjudgments when attempting to cross. In the absence of designated crosswalks, pedestrians may attempt to cross the road at unsafe locations, increasing the risk of collisions.

*5.6.3 Driver characteristics*

The rule R21 (L = 2.18) recognizes the intuitive connection between elderly drivers (>64 years) and their involvement in crashes and 'driver condition' as the primary contributing factor. Given that older drivers tend to have less cognitive and decision-making capacity, the unexpected presence of pedestrians in the D2 zone is more likely to result in pedestrian crashes. Another important relationship was identified – 'careless operation' of drivers in the D2 zone resulting in severe injury of pedestrians [R24]. This is in line with earlier research that found that driving in an unpredictable, reckless, irresponsible, negligent, or aggressive manner was linked to serious injuries to pedestrians (Behnood & Mannering, 2016).

*5.6.4 Prior Movement of Pedestrians*

Rule R23 establishes a critical link between the behavioral patterns of pedestrians and temporal factors. The combination of 'prior movement' of pedestrians as a primary contributing factor and '12 am – 6 am' hours are more likely (L = 1.59) to result in pedestrian crashes at the D2 zone. During the late-night and early morning hours, traffic volume is generally lower compared to peak hours. This can lead to drivers becoming more complacent and less vigilant while approaching an intersection (D2 zone), which may result in them not paying enough attention to the surroundings. On top of this, the 'unexpected' prior movement of pedestrians adds to the intricacy of the issues resulting in pedestrian crashes. Another rule [R38] identified the 'prior movement' of pedestrians as a possible primary contributing factor to crashes in residential settings.

### 5.7 Pattern Identification by LIC Measure for D2 zone

Table 8 provides details of rules generation for zone D2. As previously mentioned, the support parameter was progressively lowered only until the number of rules either approached about 10,000 or stopped increasing further.

**Table 8. Details of Rules Generation for Each Itemset Category (D2)**

| Itemset | Support (S) | Total rules generation with C = 0.5 |
|---|---|---|
| 3-itemset | 0.01 | 0 |
| | 0.001 | 61 |
| | 0.0001 | 170 |
| | 0.00001 | 170 |
| 4-itemset | 0.01 | 8 |
| | 0.001 | 1,903 |
| | 0.0001 | 7,467 |
| | 0.00001 | 7,467 |
| 5-itemset | 0.01 | 56 |
| | 0.001 | 14,378 |
| | 0.0013 | 9,729 |

Based on an extensive review process, 35 rules (L53 – L87) were retained for subsequent analysis (details are provided in **Table 12** in Appendix section). The following diagram shows details of the 35 rules for zone D2.

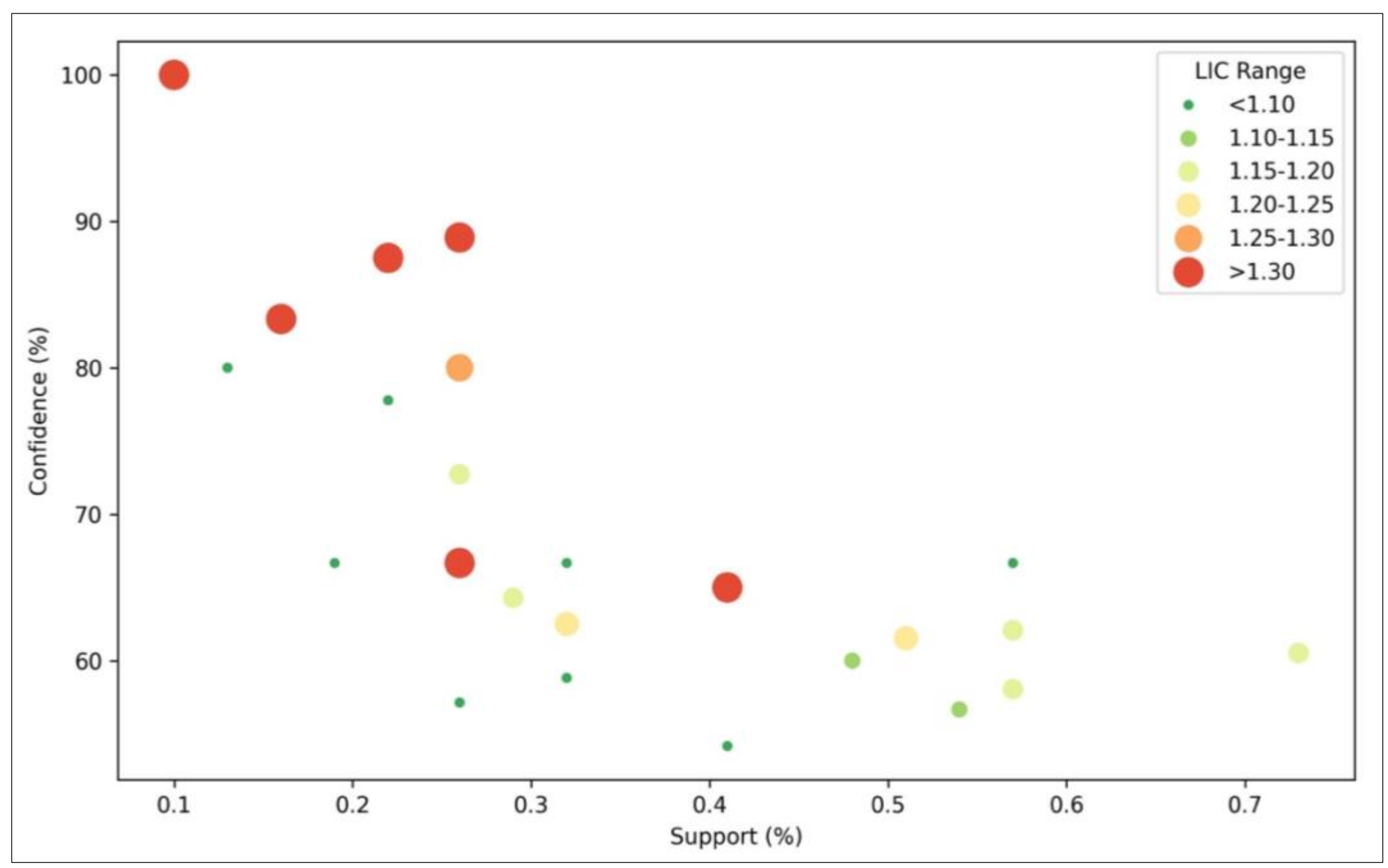

**Figure 11. Details of Association Rules (L53 – L87) for Zone D2**

The five mother rules establish foundational patterns for pedestrian crashes occurring 151 - 435 feet from intersections (D2 zone), representing the midblock transition area between intersection influence and true midblock segments. L53 (elderly drivers with driver condition as primary factor, L = 2.18) shows the strongest individual predictor, L61 (late-night prior movement crashes, L = 1.59) captures temporal-behavioral patterns, L68 (prior movement in dark streetlight residential areas, L = 1.85) reflects environmental complexity, L77 (two-way undivided roads with intoxicated pedestrians and vans/SUVs, L = 1.51) highlights impairment-vehicle interactions, and L83 (dark conditions without streetlights in residential low-speed zones, L = 1.48) demonstrates how absence of lighting infrastructure are associated with crashes in this intermediate zone where pedestrians may attempt midblock crossings away from intersection controls.

Mother rule L53's originated rules (L54-L60) reveal how elderly driver condition issues concentrate crashes in the D2 zone through specific severity and contextual factors. Fatal crashes (L54, L = 2.96, LIC = 1.36) and higher speed limits of 50-55 mph (L55, L = 2.96, LIC = 1.36) produce the strongest rule, indicating that when elderly drivers experience condition-related impairments (illness/fatigue/distraction), they are particularly likely to strike pedestrians in this midblock transition zone at higher speeds with fatal outcomes. Afternoon crashes (L56, LIC =

1.19), driver inattention (L57, LIC = 1.13), and illness/fatigue (L58, LIC = 1.09) show moderate increases, suggesting that elderly driver limitations manifest as detection failures in the D2 zone where pedestrians are less expected than at marked intersections. Mother rules L61 and L68 demonstrate temporal and environmental patterns where prior movement factors (vehicles already in motion, not turning or starting) concentrate D2 crashes. L61's originated rules show that late-night prior movement crashes intensify dramatically in residential areas (L62, L = 2.66, LIC = 1.67), with dark streetlight conditions (L64, LIC = 1.15) and two-way undivided roads (L65, LIC = 1.11) also showing elevated risk - reflecting how nighttime residential street environments create risky situation at midblock (away from intersections) where drivers maintain through-movement speeds. L68's originated rules reveal that dark streetlight conditions in residential areas with prior movement factors become most hazardous with light trucks (L69, LIC = 1.42), late-night hours (L70, LIC = 1.40), and middle-aged pedestrians (L71-L72, LIC = 1.28-1.33). This indicates that the combination of reduced visibility, vehicle size, and residential settings creates a risky situation in the D2 zone and more likelihood of pedestrian crashes where driver expectation of pedestrians is lower than at controlled intersections.

Mother rules L77 and L83 reveal how impairment and infrastructure deficiencies create D2 crash patterns. L77's originated rules (L78-L82) show that intoxicated pedestrians on two-way undivided roads with vans/SUVs experience elevated D2 risk in business/industrial areas (L78, LIC = 1.31), dark conditions without streetlights (L79, LIC = 1.23), and when pedestrian actions are the primary factor (L80, LIC = 1.21). This demonstrates that impaired pedestrians are more likely to be involved in crashes in the D2 zone where larger vehicles with poor lighting conditions are a critical factor. The absence of driver violations (L81, LIC = 1.19) in many of these crashes suggests that intoxicated pedestrians create unpredictable crossing situations that even attentive drivers cannot avoid crashes in the midblock zone. L83's originated rules (L84 - L87) reveal that dark residential streets without streetlights in low-speed zones become particularly dangerous during late-night hours (L84, lift = 2.46, LIC = 1.67) and with moderate injuries (L85, LIC = 1.30). This indicates that the absence of street lighting in residential areas creates hazardous situations in the D2 zone where pedestrians attempt crossings in total darkness, believing low speed limits provide safety, yet drivers cannot detect them until impact is unavoidable in this intermediate distance range where neither intersection controls nor midblock visibility aids exist.

### 5.8 Crashes Associated with D3 Location

To mine the association rules for case 3, the 'distance_to_intersection' variable was set to 'D3' as a 'consequent' in the model. After multiple rounds of trial and error, the minimal support values (0.1%) and confidence (55%) were selected. The program initially produced 1522 rules, many of which were repetitive. After pruning, 624 rules remained, which were sorted by lift and support values in descending order. **Table 9** lists the top 20 rules (R41-R60) for this case. In addition, the rules are graphically represented in **Figure 12**. Rules with 100% confidence, such as R41 to R45, that whenever the conditions (antecedents) of these rules are present, the outcome (consequent) is guaranteed to occur. These rules are positioned at the very top of the plot, indicating absolute reliability when specific conditions occur. Moving to the right side of the plot, rules like R51, R52, and R53 have higher support values (up to 5%) but lower confidence (around 60 - 70%). These rules apply to a larger number of crash cases, meaning they are more generalizable across the dataset.

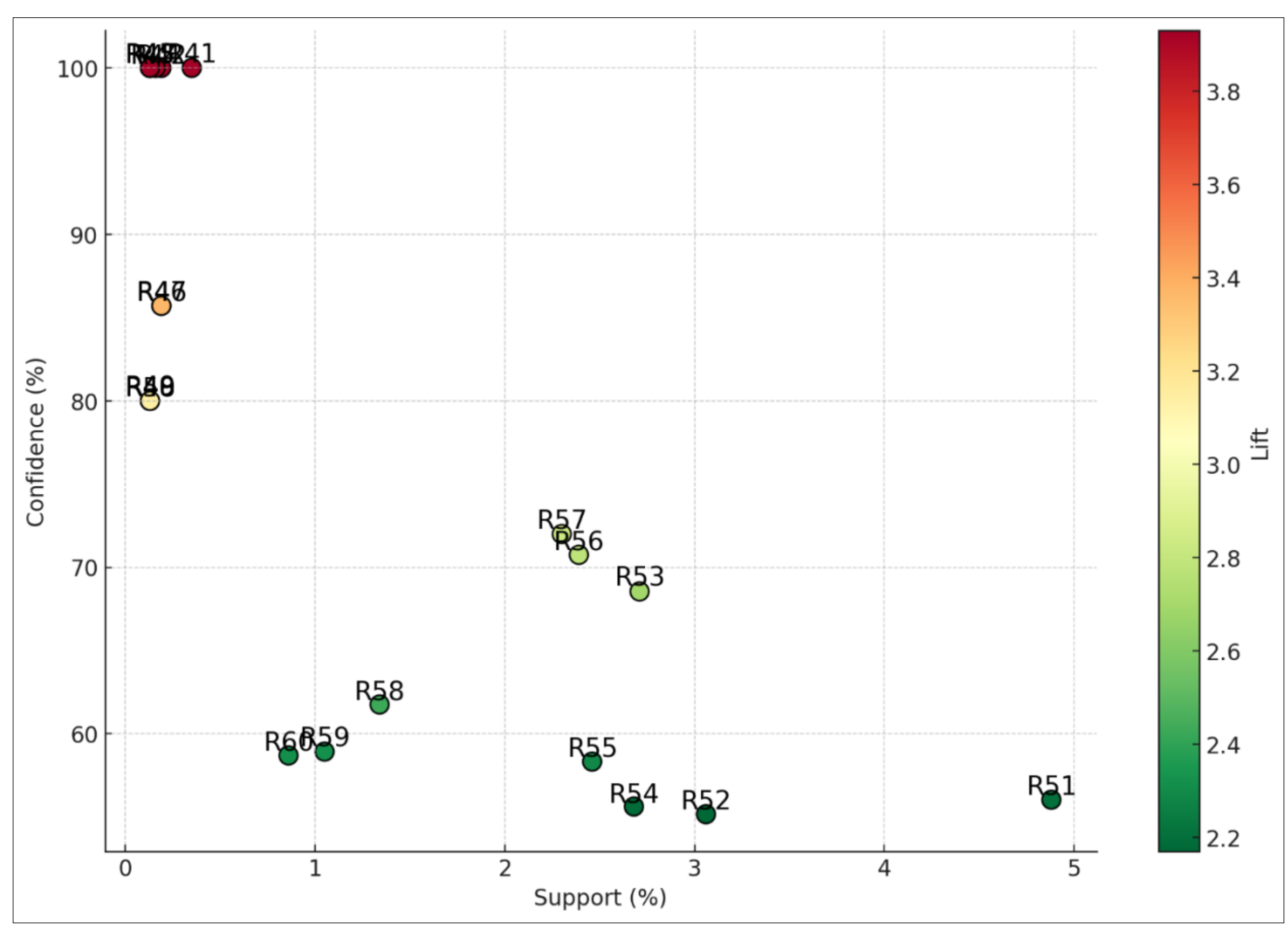

**Figure 12. Top 20 Association Rules for D3 Zone**

**Table 9. Pedestrian Crash Pattern at D3 Zone (greater than 435 ft. of an intersection)**

| **Selected top rules based on Lift value (R41-R50)** | | | | | |
|---|---|---|---|---|---|
| ID | Antecedent | S (%) | C (%) | L | # |
| R41 | {location_type=open_country, posted_speed_limit=>55} | 0.35 | 100.00 | 3.93 | 11 |
| R42 | {driver_age=15-24y, posted_speed_limit=>55} | 0.19 | 100.00 | 3.93 | 6 |
| R43 | {driver_condition=inattentive_distracted, posted_speed_limit=>55} | 0.13 | 100.00 | 3.93 | 4 |
| R44 | {lighting_condition=dusk_dawn, driver_age=25-40y, posted_speed_limit=40-45} | 0.16 | 100.00 | 3.93 | 5 |
| R45 | {ped_age=>64y,posted_speed_limit=40-45,driver_violation=careless_operation} | 0.13 | 100.00 | 3.93 | 4 |
| R46 | {primary_factor=pedestrian_condition, location_type=open_country} | 0.19 | 85.71 | 3.37 | 6 |
| R47 | {primary_factor=prior_movement, ped_age=41-64y, posted_speed_limit=50-55} | 0.19 | 85.71 | 3.37 | 6 |
| R48 | {ped_alcohol_drug=yes, posted_speed_limit=>55} | 0.13 | 80.00 | 3.14 | 4 |
| R49 | {lighting_condition=dark_no_streetlight, driver_violation=failure_to_yield} | 0.13 | 80.00 | 3.14 | 4 |
| R50 | {ped_injury=fatal, crash_hour=12pm-6pm, ped_alcohol_drug=yes} | 0.13 | 80.00 | 3.14 | 4 |
| **Selected Rules based on support value (R51-R60)** | | | | | |
| R51 | {lighting_condition=dark_no_streetlight, posted_speed_limit=50-55} | 4.88 | 56.04 | 2.20 | 153 |
| R52 | {posted_speed_limit=50-55, vehicle_type=light_truck} | 3.06 | 55.17 | 2.17 | 96 |
| R53 | {location_type=open_country, road_type=two_no_separation} | 2.71 | 68.55 | 2.69 | 85 |

| R54 | {road_type=two_no_separation, ped_age=25-40y, posted_speed_limit=50-55} | 2.68 | 55.63 | 2.19 | 84 |
|---|---|---|---|---|---|
| R55 | {ped_injury=fatal, posted_speed_limit=50-55} | 2.46 | 58.33 | 2.29 | 77 |
| R56 | {location_type=open_country, posted_speed_limit=50-55} | 2.39 | 70.75 | 2.78 | 75 |
| R57 | {lighting_condition=dark_no_streetlight, location_type=open_country} | 2.30 | 72.00 | 2.83 | 72 |
| R58 | {ped_injury=fatal, lighting_condition=dark_no_streetlight, vehicle_type=passenger_car} | 1.34 | 61.76 | 2.43 | 42 |
| R59 | {lighting_condition=dark_no_streetlight, ped_age=15-24y, vehicle_type=light_truck} | 1.05 | 58.93 | 2.32 | 33 |
| R60 | {ped_injury=fatal, lighting_condition=dark_no_streetlight, ped_alcohol_drug=yes} | 0.86 | 58.70 | 2.31 | 27 |

*Note: S = Support, C = Confidence, L = lift, L.H.S = Left-Hand-Side of association rules, # indicates number of occurrences in the database

Some of the major crash patterns are discussed under the following subheadings:

*5.8.1. High-speed roadways (>55 mph, 50-55 mph)*

Pedestrian crashes occurring in the D3 zone were found to be characterized by high posted speed limits (50-55 mph or >55 mph). Some of the other factors (location type, driver/pedestrian characteristics, and environmental condition) were found to be associated with such high-speed settings. For example, young drivers were more likely (L = 3.93) to be involved in pedestrian crashes on roadways having a posted speed limit of 55 mph or higher [R42]. In addition, the 'inattentive/distracted' condition of drivers was also found to be associated with pedestrian crashes on such high-speed roadways [R43]. Note that the confidence level for each of these association rules is 100%, indicating that such a critical interaction of variable attributes is unique to D3 zones.

Pedestrian crashes were also found to be associated with 'open country' locations and high-speed settings [R41, and R56]. Rule R56 has a support value of 2.39%, suggesting that there were 75 instances (2.39% multiplied by 3,135) in which pedestrians were involved in crashes in open country locations and the posted speed roadway was 50-55 mph. Note that this critical combination is 2.78 times more likely to occur in the D3 zone. Previous research has also identified the increased likelihood of pedestrian crashes (fatal or severe injuries) in open country locations (Das et al., 2020; A. Hossain, Sun, Mahmud Zafri, et al., 2023; A. Hossain, Sun, Thapa, et al., 2023b; S. Islam & Jones, 2014a). Drivers in open country locations may not expect to encounter pedestrians frequently. This lack of awareness can lead to reduced caution and a higher likelihood of not noticing pedestrians in time. Dark lighting conditions are expected to add to the complexity of pedestrian crashes in high-speed settings, identified by rule R51 (L = 2.20). Some other factors contributing to pedestrian crashes on high-speed settings were – [R47] prior movement of pedestrians of age group 41-64 years (L = 3.37), [R48] pedestrian alcohol impairment (L = 3.14), [R52] involvement with light-truck vehicle type (L = 2.17), [R54] involving pedestrians aged 25-40 years on two-way roadways with no physical separation, and [R55] resulting in pedestrian fatalities (L = 2.29).

*5.8.2 Dark-no-streetlight condition*

Previous studies have clearly demonstrated the critical connection between poor lighting conditions and pedestrian crashes (Haleem et al., 2015; Sullivan & Flannagan, 2007, 2011; Sun et al., 2019). Consistent with prior research, this investigation also identified 'dark-no-streetlight' as a potential contributor to pedestrian crashes occurring in the D3 zone [R49, R51, R57, R58, R59, and R60]. Rule 49 states that pedestrians are more likely to be involved in crashes in the 'dark-no-streetlight' condition due to the driver's 'fails to yield' (L = 3.14). In dark lighting conditions, the visibility of pedestrians can be significantly reduced, making it more challenging for drivers to detect them. Consequently, the driver's reaction time to unexpected situations could be slower, making it harder for them to stop or yield in time when a pedestrian appears suddenly in the D3 zone.

Two of the rules identified complex interactions of multiple factors contributing to pedestrian fatal injury in dark-no-streetlight conditions [R58, R60]. According to rule R60, alcohol-impaired pedestrians were more likely to be involved in fatal crashes in dark-no-streetlight conditions (L = 2.31). When pedestrians consume alcohol and attempt to crossroads or walk near traffic in poorly lit areas, their impaired judgment, decreased coordination, and reduced reaction times significantly increase the risk of crashes. From the driver's perspective, alcohol consumption can lead to unpredictable behavior, making it challenging for drivers to anticipate the actions of impaired pedestrians. This unpredictability increases the likelihood of fatal crashes in the D3 zone.

### *5.8.3 Pedestrian characteristics*

Crashes occurring in the D3 zone were also characterized by behavioral and demographic traits of pedestrians and identified by a few of the rules (R45, R46, R47, R48, R50, R54, and R60). Rule R45 states that older pedestrians (>64 years) were more likely to be involved in crashes due to the 'careless operation' of drivers on roadways having a posted speed limit of 40-45 mph (L = 3.93). From a general context, older pedestrians are more willing to wait at the crosswalks and obey the traffic rules (high safety perception). However, with a lack of crosswalks on high-speed roadways, older pedestrians may be more likely to cross the road at non-intersection locations due to factors like a desire to take more direct routes or difficulties with walking longer distances to reach crosswalks. Additionally, older pedestrians may walk at a slower speed, which can increase their exposure time to traffic while crossing the road or navigating non-intersection locations. Pedestrian alcohol impairment was identified as a contributor to pedestrian crashes occurring in the D3 zone [R48, R50, and R60]. In addition, the pedestrian condition was identified as a factor contributing to pedestrian crashes in open country locations [R46].

## 5.9 Pattern Identification by LIC Measure for D3 zone

**Table 10** provides details of rules generation for zone D3. As previously mentioned, the support parameter was progressively lowered only until the number of rules either approached about 10,000 or stopped increasing further.

**Table 10. Details of Rules Generation for Each Itemset Category (D3)**

| Itemset | Support (S) | Total rules generation with C = 0.5 |
|---|---|---|
| 2-itemset | 0.01 | 1 |
| | 0.001 | 2 |
| | 0.0001 | 2 |
| 3-itemset | 0.01 | 39 |
| | 0.001 | 153 |
| | 0.0001 | 264 |
| | 0.00001 | 264 |
| 4-itemset | 0.01 | 180 |
| | 0.001 | 2054 |
| | 0.0001 | 7280 |
| | 0.00001 | 7280 |
| 5-itemset | 0.01 | 250 |

|  | 0.001 | 12100 |
|---|---|---|
|  | 0.0013 | 8720 |

Based on an extensive review process, 37 rules (L88 – L124) were retained for subsequent analysis (details are provided in **Table 13** in Appendix section). The following diagram shows details of the 37 rules for zone D3.

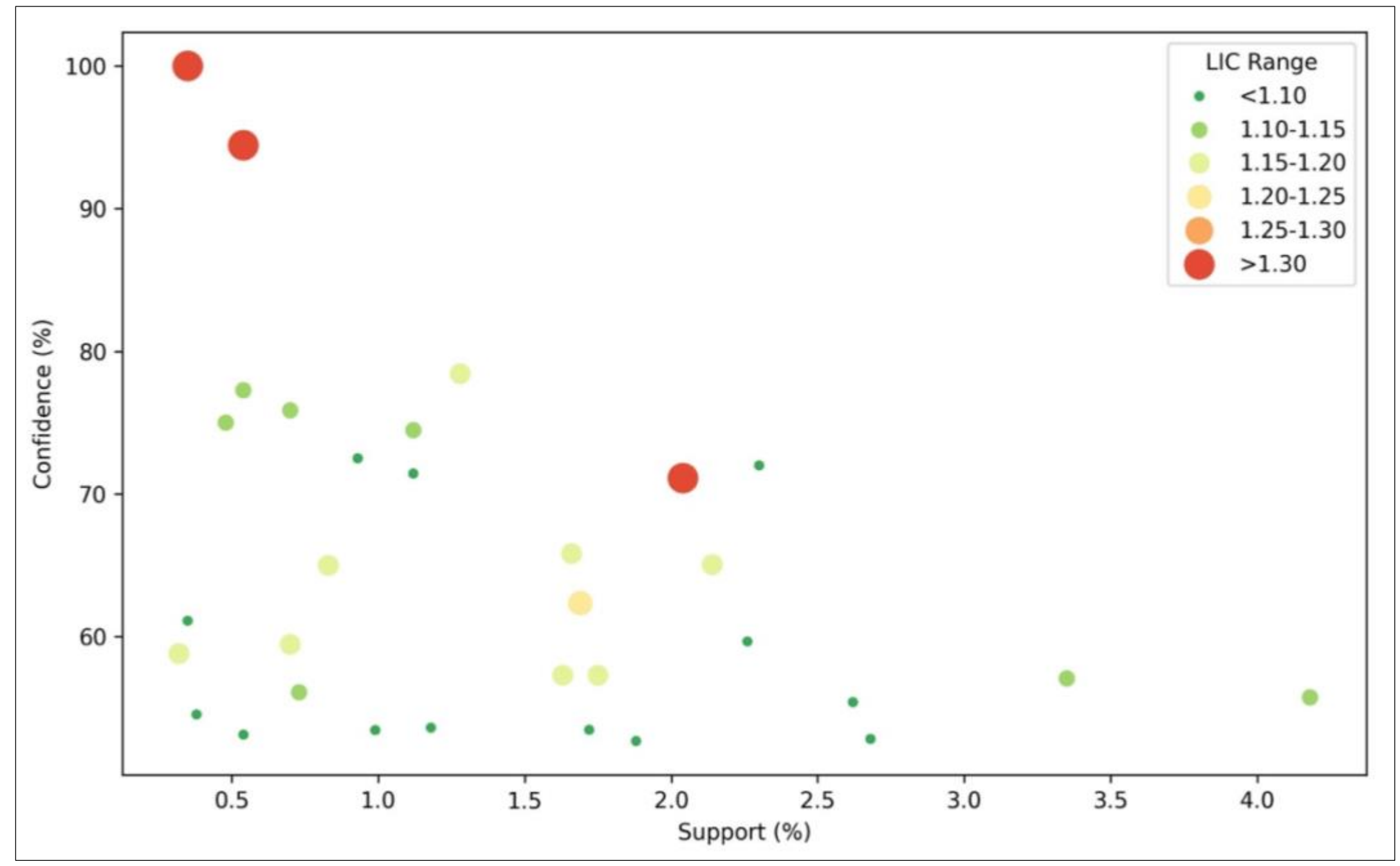

**Figure 13. Details of Association Rules (L88 – L124) for zone D3**

The four mother rules establish foundational patterns for pedestrian crashes occurring more than 435 feet from intersections (D3 zone), representing segments where intersection influence is minimal, and pedestrians often cross at unmarked locations. L88 (open country locations, L = 2.67, S = 3.57%) shows the strongest individual predictor, L99 (dark conditions without streetlights at 50-55 mph speeds, L = 2.20, S = 4.88%) captures the environmental-speed interaction, L105 (two-way undivided roads at 50-55 mph, L = 2, S = 6.54%) reflects infrastructure characteristics, and L114 (50-55 mph speeds with non-violating drivers, L = 1.96, S = 4.94%) demonstrates how higher-speed environments dominate this distant zone. These mother rules collectively emphasize that D3 crashes occur in rural or semi-rural settings where higher speeds, absent lighting infrastructure, and lack of pedestrian facilities create hazardous conditions with absence of intersection controls.

Mother rule L88's originated rules (L89-L98) reveal how open country locations concentrate D3 crashes through speed and environmental factors. The addition of speeds exceeding 55 mph (L89, L = 3.93, LIC = 1.47) and other contributing factors (L90, LIC = 1.39) produce the strongest rules, indicating that rural open country settings with high-speed roadways create the most extreme segment crash patterns where pedestrians - often walking along roadways due to absent sidewalks - are struck far from any intersection. Light trucks (L91, LIC = 1.16), morning hours (L92, LIC = 1.14), and middle-aged pedestrians (L95, LIC = 1.10) show moderate increases, reflecting how open country environments attract diverse scenarios where pedestrian presence is unexpected and vehicle speeds remain high throughout road segments. The prevalence of dark conditions without streetlights (L97, LIC = 1.06) confirms that rural roadways lacking illumination result in crashes in the D3 zone where pedestrians crossing or walking along roads become invisible until the final moments before impact.

Mother rules L99 and L105 demonstrate how speed and infrastructure interact to create D3 patterns. L99's originated rules show that dark, unlit conditions at 50-55 mph become most deadly with fatal outcomes (L100, LIC = 1.17), light trucks (L101, LIC = 1.16), morning crashes (L102, LIC = 1.16), and pedestrian actions as primary factors (L104, LIC = 1.06). This indicates that this high-speed lighting combination creates midblock zones where detection distances are insufficient even for attentive drivers, particularly with larger vehicles requiring longer stopping distances. L105's origin rules reveal that two-way undivided roads at 50-55 mph intensify D3 risk dramatically in open country settings (L106, L = 2.79, LIC = 1.40), with fatal crashes (L107, LIC = 1.23), pedestrian condition factors (L108, LIC = 1.16), and dark conditions (L110, LIC = 1.10) showing elevated associations. This pattern reflects how undivided high-speed roadways in rural areas create situations where pedestrians must cross without refuge islands or marked crosswalks, facing vehicles traveling at speeds that provide minimal reaction time for unexpected pedestrian presence far from intersections.

Mother rule L114 and its originating rules (L115 - L124) reveal a critical paradox: crashes occurring at 50-55 mph with non-violating drivers show strong D3 associations, indicating systemic infrastructure failures rather than driver behavior issues. Open country locations (L115, lift = 2.96, LIC = 1.51) produce the strongest rules, followed by normal pedestrian condition (L116, LIC = 1.19), fatal outcomes (L117, LIC = 1.15), middle-aged pedestrians (L118, LIC = 1.15), and dark unlit conditions (L119, LIC = 1.14). These patterns demonstrate that the D3 zone represents a fundamental mismatch between roadway design and pedestrian presence. The high-speed roads designed for vehicular throughput are being used by pedestrians who lack safe crossing infrastructure, resulting in crashes where drivers commit no violations yet still strike pedestrians because the road geometry, speed environment, and absence of pedestrian facilities make crashes unavoidable.

## 6. CONCLUSIONS

This study investigates non-intersection crashes involving pedestrians using a crash database (2017-2021) collected from Louisiana State. As the risk of pedestrian crashes tends to vary with distance from the intersection, the research team utilized a unique framework, 'distance to intersection' to capture the differences in crash patterns at non-intersection locations. The collected 3,135 pedestrian crashes at non-intersection locations during the study period were subdivided into three zones - D1 (area within intersection influence) designates crashes occurring within 150 ft. of an intersection (1,277 crashes), D2 (transitional area) designates crashes occurring within 151 ft. to 435 ft. of an intersection (1,060 crashes) and D3 (area out of intersection influence)designates crashes occurring at 435 ft. or higher from an intersection (798 crashes). To explore the complex interaction of multiple factors, an intuitive data mining technique, Association Rules Mining, was used. A total of the top 60 interesting association rules (20 for each zone) providing key information on pedestrian crashes at non-intersection locations were identified by the algorithm based on two contexts: a) low support but high confidence, b) high support but low confidence.

Focusing on the area within intersection influence, pedestrian characteristics played a critical role in crashes near intersections. For instance, pedestrians classified as 'inattentive/distracted' were found to have a higher likelihood of being involved in crashes, particularly on one-way roads within the D1 zone. Moreover, pedestrian violations and their actions were identified as critical factors contributing to crashes in this area in combination with specific criteria (business with mixed residential area, pedestrian age <15 years, one-way road). From the driver's perspective, instances of 'failure-to-yield' were noted as significant contributors to pedestrian crashes within the D1 zone. Additionally, the study highlighted the involvement of older drivers (aged over 64 years) in pedestrian crashes in this zone.

Pedestrian crashes occurring in the transitional area are mainly characterized by high posted speed in the range of 40-45 mph. A few other factors associated with this specific setting include dark lighting conditions, pedestrian alcohol impairment, Van/SUV type vehicle, and driver age group between 41-64 years. Pedestrians are more likely to be involved in crashes on two-way roads with no physical separation in the D2 zone. Driver characteristics also play a critical role in crashes, such as the combination of elderly drivers (>64 years) and 'driver condition' as a primary contributing factor. A combination of prior movement of pedestrians and '12 am – 6 am' is identified as a contributing factor to pedestrian crashes in the D2 zone. Pedestrian crashes occurring in the 'area out of intersection influence' are primarily associated with high posted speeds, particularly exceeding 50 mph, potentially suggesting that the higher the distance from the intersection, the more likely that the crash is expected to occur on high-speed settings. The two other critical factors associated with pedestrian crashes in the D3 zone include 'open country' location and dark-no-streetlight conditions. Note that pedestrian fatal injury is identified in a few of the association rules, potentially suggesting the impact of high-speed vehicles in combination with other factors, including dark lighting conditions, pedestrian alcohol impairment, passenger car, 12 pm – 6 pm.

Reducing pedestrian crashes at non-intersections requires a comprehensive approach that involves a combination of engineering, education, and enforcement strategies. Based on the current research findings, some of the countermeasures are recommended to address the crash patterns and reduce the number of pedestrian crashes at non-intersection locations. According to the analysis, 50% of non-intersection pedestrian crashes occurred within 198 ft. of the intersection. First of all, pedestrians must be aware of such unsafe street crossing behavior at close distance from the intersection. Promoting awareness of safe crossing strategies through educational campaigns can play a vital role. To reduce crossing distance for pedestrians, some of the recommended countermeasures are curb extension, crossing islands, waiting areas, and road diet strategies (Kang, 2019; Organization, 2023; Schneider et al., 2018; Scott Suderman et al., n.d.).

Zone D1 is characterized by several factors related to pedestrians, including age (<15 years, 15-24 years), inattentive/distracted condition, pedestrian actions, and pedestrian violation. The co-occurrence of '<15 years' old pedestrians and 'inattentive/distracted' condition can possibly be attributed to digital distraction using smartphones (Gruden et al., 2021). This can be addressed through public awareness campaigns targeting both young pedestrians and their parents about the risks associated with distracted walking. Another critical factor is the driver's violation and failure to yield to pedestrians at the D1 zone. An effective approach to solve this critical safety concern at the D1 zone is to restrict the number of right-turn-red maneuvers in areas with high pedestrian activity, and to limit left and right-turn speeds to 20 mph and 15 mph, respectively (*FHWA Course on Bicycle and Pedestrian Transportation - Safety, Federal Highway Administration*, 2006).

Built environment characteristics such as high posted speed limit (40-45 mph), two-way roads with no physical separation, and dark lighting conditions (with or without streetlights) play a critical role in pedestrian safety at the D2 zone. Reducing approach speeds at high-speed intersections can be accomplished effectively by installing transverse rumble strips (Yang et al., 2016). To improve drivers' yielding behavior, innovative ITS-based solutions are also recommended, including LED pavement lights and Variable Message Signs (VMS) (Hussain et al., 2023). Rectangular Rapid-Flash Beacons (RRFBs) and Pedestrian Hybrid Beacons (PHB) at mid-block locations are other proven safety countermeasures to increase driver yielding behavior (Brewer et al., 2015; Fitzpatrick et al., 2016; Porter et al., 2016).

Pedestrian crashes associated with zone D3 are characterized by high posted speed limit (50-55 mph, >55 mph), dark-no-streetlight conditions, open country location, and pedestrian alcohol involvement. Visibility is a significant factor contributing to pedestrian crashes at night, especially in high-speed settings (A. Hossain et al., 2022; A. Hossain, Sun, Mahmud Zafri, et al., 2023; A. Hossain, Sun, Shahrier, et al., 2023c; A. Hossain, Sun, Thapa, et al., 2023b). From pedestrians' perspective, they are usually unaware of their nighttime conspicuity problem (R. Tyrrell et al., 2004), and are involved in risky behaviors including crossing and walking with or against traffic. Pedestrians must be aware of their conspicuity problem at night. In addition, the dangers of 'Walking under the influence' (WUI) of alcohol must be addressed to improve pedestrian safety at the D3 zone.

The LIC-based analysis reveals distinct crash patterns corresponding to intersection proximity and infrastructure context. D1 zone crashes are characterized by infrastructure-behavioral interactions where one-way streets, larger vehicles (vans/SUVs), and attention deficits by either road user concentrate crashes at critical decision-making points where turning movements, pedestrian crossings, and traffic controls converge. Environmental complexity in business/mixed residential areas under dark streetlight conditions amplifies these patterns, with the highest lift value increases occurring when geometric features combine with inattention or careless operation. D2 zone crashes represent a transitional midblock area dominated by elderly driver condition issues, late-night prior movement crashes in residential settings, and impaired pedestrian behavior on undivided roadways, reflecting how this intermediate distance loses the protective effect of intersection controls while encouraging riskier midblock crossing attempts, particularly in dark conditions where streetlight coverage becomes inconsistent and driver expectation of pedestrians diminishes. D3 zone crashes demonstrate systemic infrastructure failures in rural environments where open country locations, high speeds (50-55+ mph), two-way undivided roads, and complete absence of street lighting create lethal conditions; the prevalence of non-violating drivers in fatal D3 crashes indicates that these incidents result not from driver error but from fundamental design inadequacies where high-speed roadways lack pedestrian facilities, forcing pedestrians into unavoidable conflicts with vehicles traveling at high speeds.

The uniqueness of this research lies in its novel data mining approach to address non-intersection pedestrian crashes using 'distance to intersection' as a measure to identify the associated patterns. The major limitation of traditional statistical methodology compared to data mining techniques like ARM is its reliance on predefined hypotheses and assumptions, which can restrict the scope of crash data analysis and the nature of insights it provides. Traditional statistical methods often require assumptions of normality, independence, and linearity, and are typically used for hypothesis testing, which can limit their applicability in complex, real-world crash scenarios where multiple factors act concurrently to cause a crash (Rella Riccardi et al., 2022). In contrast, ARM is designed to handle complex interactions of multiple crash contributing factors without the need for prior hypotheses, making it highly effective for discovering insightful patterns and associations (Ait-Mlouk, Gharnati, et al., 2017; Feng et al., 2020; Hong et al., 2020; Sivasankaran et al., 2020). In addition, conventional statistical models typically focus on assessing the average effects of risk factors, often overlooking the presence of subgroups with distinct risk patterns (Haghighi et al., 2016). In contrast, rule-based analysis methods can uncover these subgroups with heterogeneous risk profiles without requiring any assumptions about the subgroups or the analytical method. These methods can characterize the risk patterns of specific subsets of the data by accounting not only for interactions between risk factors but also for the specific ranges of these factors.

### 6.1 Practical Applications

The practical application of assessing intersection proximity in pedestrian crashes through this data mining approach is to inform targeted, zone-specific safety interventions. By utilizing ARM to identify and categorize critical crash factors across different proximity zones (D1, D2, D3), transportation agencies can prioritize and customize safety strategies tailored to each zone's unique risk profile (see **Figure 14** for more details).

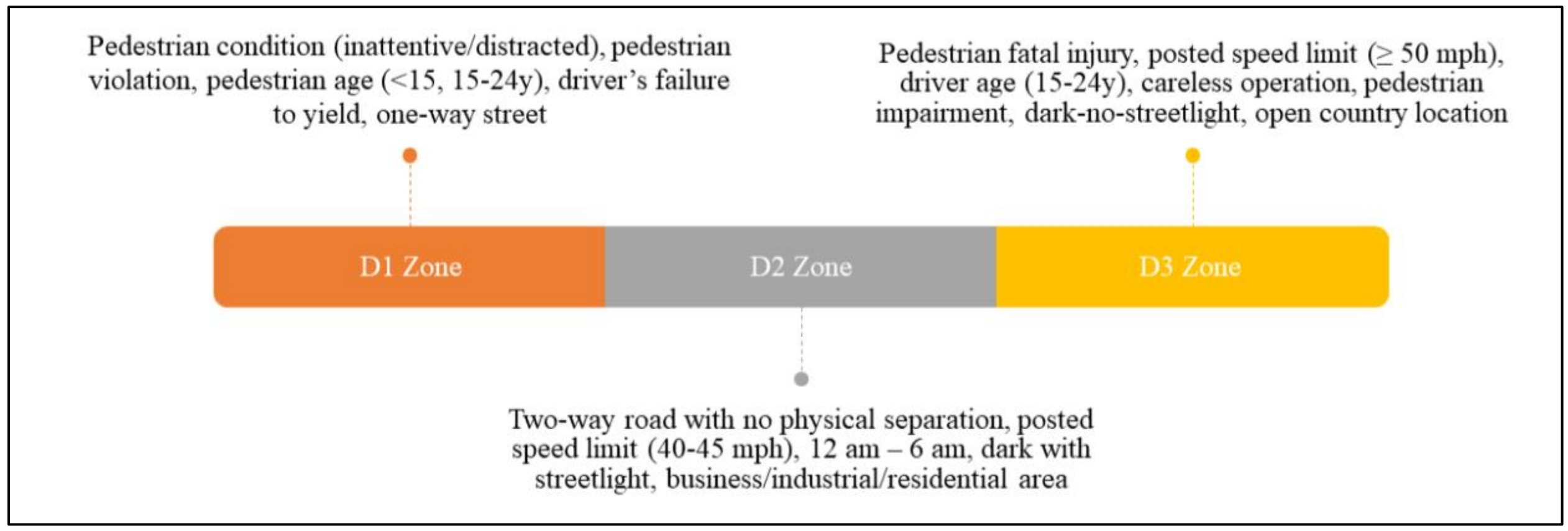

**Figure 14. Identified critical crash risk factors at D1, D2, and D3 zone**

The findings suggest targeted safety measures for each proximity zone to intersections to reduce pedestrian crashes. In the D1 zone, close to intersections, interventions should address pedestrian inattention and violations, enhance driver awareness to yield, and manage risks for younger pedestrians and one-way streets by implementing improved crosswalk visibility, pedestrian warning systems, and traffic calming measures. For the D2 zone, located mid-range from intersections, strategies should focus on mitigating moderate-speed, nighttime risks on two-way streets in business, industrial, or residential areas; practical applications include upgrading street lighting, reviewing speed limits, and introducing physical separation measures on two-way roads to enhance nighttime pedestrian safety. In the D3 zone, farther from intersections and characterized by high-speed, rural settings with severe crash outcomes, robust measures such as installing pedestrian overpasses or underpasses, enhancing signage and warnings for high-speed zones, and enforcing regulations against impaired and careless driving are critical to protecting pedestrians. These zone-specific interventions offer actionable insights to guide transportation professionals in designing effective, location-based pedestrian safety enhancements.

From a practical and policy perspective, the lack of a significant influence of area setting (e.g., urban or rural) on distance-related pedestrian crash patterns indicates that safety interventions focused on reducing pedestrian crashes near intersections may be broadly effective across different area classifications. Consequently, countermeasures targeting distance-related risk factors such as optimized crosswalk placement, improved sight distance, and signal timing adjustments can be implemented without substantial modification based on area setting. This finding simplifies the development and deployment of pedestrian safety strategies and supports their applicability across diverse roadway environments.

### 6.2 Study Limitations

This study is not without limitations. Several built environment characteristics (crossing distance, presence of bus stops, turning lanes, presence of median, driveway density, and traffic control) and exposure variables (e.g., pedestrian volume) are expected to impact the likelihood of pedestrian crashes at non-intersection locations. Specifically, the roadway characteristics leading to the intersection are expected to vary by number of lanes (e.g., exclusive left turn lanes), roadway width, cross-section, and control type and affect crashes occurring at the D1 zone (within 150 ft. of the intersection). The inclusion of such built environment and exposure variables is thus recommended as a future study.

Another limitation of this research is that the research team relied on the latitude-longitude information reported by the police at the crash location, and using this information, the distance to the intersection was measured. The accuracy of the latitude-longitude data reported by the police may vary depending on multiple circumstances. Factors such as human error in recording or transcription, limitations of the reporting system, or technical issues could affect the precision of the data. Verification of such data input issues is beyond the scope of current research. One limitation of this study is the potential inaccuracy in the recorded crash locations, particularly in cases where the variable 'CRASH_TB_intersection' indicates that a crash did not occur at the intersection, yet the measured distance to the intersection is zero. This discrepancy might arise from the rounding of latitude and longitude coordinates, which can cause crashes that occur very close to the intersection to appear as if they happened within the intersection. By excluding these cases, this study aimed to ensure the clarity and consistency of our analysis. However, this decision may have inadvertently disregarded crashes that were near enough to the intersection to be relevant but were not recorded as such due to minor differences in geographic data. This limitation highlights the challenge of accurately capturing crash locations and suggests that future research should consider methods to better account for these near-intersection crashes, possibly by incorporating a buffer zone around intersections or addressing the nuances of geographic data rounding in crash analysis. Another important idea is to identify the exposure context (e.g., hazard rates, relative risk) according to distance to the intersection, which can provide critical insights. Future research should focus on spatial analysis of crash data to map hazard rates at varying distances from intersections, incorporating roadway and environmental factors, and conducting time-dependent studies to capture temporal risk variations.

## ACKNOWLEDGMENTS

The research team would like to acknowledge the support of the Louisiana Department of Transportation Development (LaDOTD) for supplying the database used in this study. The authors would also like to acknowledge the support of Md. Fazle Rabbi and Shahrin Islam for their support in the Google Street View step data collection process.

## FUNDING

The authors received no funding for conducting this research.

## CONFLICT OF INTEREST

The author(s) declare(s) that there is no conflict of interest regarding the publication of this paper.

# Appendix

## Table 11. Crash Pattern Identification by LIC for zone D1

| ID | rules | S (%) | C (%) | L | # | LIC |
|---|---|---|---|---|---|---|
| L1 | {road_type=one_way} => {distance=D1} | 4.18 | 67.18 | 1.65 | 131 | N/A |
| L2 | {road_type=one_way, ped_condition=inattentive_distracted} => {distance=D1} | 0.73 | 85.19 | 2.09 | 23 | 1.27 |
| L3 | {ped_injury=PDO, road_type=one_way} => {distance=D1} | 0.41 | 76.47 | 1.88 | 13 | 1.14 |
| L4 | {road_type=one_way, ped_age=25-40y} => {distance=D1} | 1.40 | 73.33 | 1.80 | 44 | 1.09 |
| L5 | {crash_hour=12pm-6pm, road_type=one_way} => {distance=D1} | 1.21 | 73.08 | 1.79 | 38 | 1.09 |
| L6 | {road_type=one_way, posted_speed_limit=<30} => {distance=D1} | 2.14 | 72.83 | 1.79 | 67 | 1.08 |
| L7 | {crash_hour=12am-6am, road_type=one_way} => {distance=D1} | 0.57 | 72.00 | 1.77 | 18 | 1.07 |
| L8 | {primary_factor=pedestrian_actions, road_type=one_way} => {distance=D1} | 1.05 | 71.74 | 1.76 | 33 | 1.07 |
| L9 | {ped_injury=complaint, road_type=one_way} => {distance=D1} | 1.66 | 71.23 | 1.75 | 52 | 1.06 |
| L10 | {ped_injury=moderate, road_type=one_way} => {distance=D1} | 1.63 | 70.83 | 1.74 | 51 | 1.05 |
| L11 | {posted_speed_limit=<30, vehicle_type=Van_SUV} => {distance=D1} | 4.05 | 56.19 | 1.38 | 127 | N/A |
| L12 | {road_type=one_way, posted_speed_limit=<30,vehicle_type=Van_SUV} => {distance=D1} | 0.73 | 74.19 | 1.82 | 23 | 1.32 |
| L13 | {road_type=two_physical_separation, posted_speed_limit=<30,vehicle_type=Van_SUV} => {distance=D1} | 0.35 | 73.33 | 1.80 | 11 | 1.3 |
| L14 | {location_type=business_mixed_residential,posted_speed_limit=<30,vehicle_type=Van_SUV} => {distance=D1} | 0.99 | 64.58 | 1.59 | 31 | 1.15 |
| L15 | {crash_hour=12pm-6pm,posted_speed_limit=<30,vehicle_type=Van_SUV} => {distance=D1} | 1.50 | 64.38 | 1.58 | 47 | 1.15 |
| L16 | {ped_age=<15y,posted_speed_limit=<30,vehicle_type=Van_SUV} => {distance=D1} | 0.99 | 62.00 | 1.52 | 31 | 1.1 |
| L17 | {ped_age=>64y,posted_speed_limit=<30,vehicle_type=Van_SUV} => {distance=D1} | 0.35 | 61.11 | 1.50 | 11 | 1.09 |
| L18 | {ped_injury=moderate,posted_speed_limit=<30,vehicle_type=Van_SUV} => {distance=D1} | 1.85 | 60.42 | 1.48 | 58 | 1.08 |
| L19 | {primary_factor=prior_movement,posted_speed_limit=<30,vehicle_type=Van_SUV} => {distance=D1} | 0.61 | 59.38 | 1.46 | 19 | 1.06 |
| L20 | {driver_age=25-40y,posted_speed_limit=<30,vehicle_type=Van_SUV} => {distance=D1} | 1.12 | 59.32 | 1.46 | 35 | 1.06 |
| L21 | {lighting_condition=daylight,posted_speed_limit=<30,vehicle_type=Van_SUV} => {distance=D1} | 2.55 | 59.26 | 1.45 | 80 | 1.05 |
| L22 | {driver_condition=normal,posted_speed_limit=<30,vehicle_type=Van_SUV} => {distance=D1} | 2.04 | 59.26 | 1.45 | 64 | 1.05 |
| L23 | {lighting_condition=dark_with_streetlight,location_type=business_mixed_residential} => {distance=D1} | 5.74 | 53.89 | 1.32 | 180 | N/A |
| L24 | {lighting_condition=dark_with_streetlight,location_type=business_mixed_residential,driver_violation=careless_operation} => {distance=D1} | 0.80 | 73.53 | 1.81 | 25 | 1.36 |
| L25 | {ped_injury=PDO,lighting_condition=dark_with_streetlight,location_type=business_mixed_residential} => {distance=D1} | 0.35 | 73.33 | 1.80 | 11 | 1.36 |
| L26 | {lighting_condition=dark_with_streetlight,location_type=business_mixed_residential,ped_condition=inattentive_distracted} => {distance=D1} | 1.40 | 69.84 | 1.71 | 44 | 1.30 |
| L27 | {lighting_condition=dark_with_streetlight,location_type=business_mixed_residential,road_type=one_way} => {distance=D1} | 0.70 | 68.75 | 1.69 | 22 | 1.28 |
| L28 | {lighting_condition=dark_with_streetlight,location_type=business_mixed_residential,posted_speed_limit=<30} => {distance=D1} | 0.93 | 63.04 | 1.55 | 29 | 1.17 |
| L29 | {lighting_condition=dark_with_streetlight,location_type=business_mixed_residential,driver_condition=inattentive_distracted} => {distance=D1} | 0.64 | 62.50 | 1.53 | 20 | 1.16 |
| L30 | {primary_factor=prior_movement,lighting_condition=dark_with_streetlight,location_type=business_mixed_residential} => {distance=D1} | 0.89 | 62.22 | 1.53 | 28 | 1.15 |
| L31 | {lighting_condition=dark_with_streetlight,location_type=business_mixed_residential,ped_age=<15y} => {distance=D1} | 0.35 | 61.11 | 1.50 | 11 | 1.13 |
| L32 | {lighting_condition=dark_with_streetlight,location_type=business_mixed_residential,driver_condition=alcohol_drug} => {distance=D1} | 0.45 | 60.87 | 1.49 | 14 | 1.13 |
| L33 | {lighting_condition=dark_with_streetlight,location_type=business_mixed_residential,driver_age=unknown} => {distance=D1} | 1.50 | 58.75 | 1.44 | 47 | 1.09 |
| L34 | {lighting_condition=dark_with_streetlight,location_type=business_mixed_residential,ped_alcohol_drug=no} => {distance=D1} | 3.00 | 58.39 | 1.43 | 94 | 1.08 |
| L35 | {lighting_condition=dark_with_streetlight,location_type=business_mixed_residential,posted_speed_limit=unknown} => {distance=D1} | 0.51 | 57.14 | 1.40 | 16 | 1.06 |
| L36 | {lighting_condition=dark_with_streetlight,location_type=business_mixed_residential,driver_age=15-24y} => {distance=D1} | 0.77 | 57.14 | 1.40 | 24 | 1.06 |
| L37 | {ped_age=<15y,driver_violation=no_violations} => {distance=D1} | 4.21 | 50.00 | 1.23 | 132 | N/A |
| L38 | {crash_hour=6am-12pm,ped_age=<15y,driver_violation=no_violations} => {distance=D1} | 0.51 | 61.54 | 1.51 | 16 | 1.23 |
| L39 | {lighting_condition=dark_with_streetlight,ped_age=<15y,driver_violation=no_violations} => {distance=D1} | 0.83 | 60.47 | 1.48 | 26 | 1.21 |
| L40 | {location_type=business_mixed_residential,ped_age=<15y,driver_violation=no_violations} => {distance=D1} | 1.31 | 59.42 | 1.46 | 41 | 1.19 |

| L41 | {ped_alcohol_drug=unknown,ped_age=<15y,driver_violation=no_violations} => {distance=D1} | 0.61 | 52.78 | 1.30 | 19 | 1.06 |
|---|---|---|---|---|---|---|
| L42 | {lighting_condition=dark_with_streetlight,location_type=business_mixed_residential,road_type=two_no_separation} => {distance=D1} | 3.48 | 54.50 | 1.34 | 109 | N/A |
| L43 | {lighting_condition=dark_with_streetlight,location_type=business_mixed_residential,road_type=two_no_separation, driver_age=15-24y} => {distance=D1} | 0.51 | 72.73 | 1.79 | 16 | 1.33 |
| L44 | {primary_factor=prior_movement,lighting_condition=dark_with_streetlight,location_type=business_mixed_residential, road_type=two_no_separation} => {distance=D1} | 0.54 | 68.00 | 1.67 | 17 | 1.25 |
| L45 | {lighting_condition=dark_with_streetlight,location_type=business_mixed_residential,road_type=two_no_separation, driver_condition=alcohol_drug} => {distance=D1} | 0.38 | 66.67 | 1.64 | 12 | 1.22 |
| L46 | {lighting_condition=dark_with_streetlight,location_type=business_mixed_residential,road_type=two_no_separation, driver_violation=careless_operation} => {distance=D1} | 0.38 | 66.67 | 1.64 | 12 | 1.22 |
| L47 | {lighting_condition=dark_with_streetlight,location_type=business_mixed_residential,road_type=two_no_separation, ped_condition=inattentive_distracted} => {distance=D1} | 0.96 | 66.67 | 1.64 | 30 | 1.22 |
| L48 | {lighting_condition=dark_with_streetlight,location_type=business_mixed_residential,road_type=two_no_separation, driver_condition=inattentive_distracted} => {distance=D1} | 0.38 | 63.16 | 1.55 | 12 | 1.16 |
| L49 | {lighting_condition=dark_with_streetlight,location_type=business_mixed_residential,road_type=two_no_separation, vehicle_type=passenger_car} => {distance=D1} | 1.59 | 60.98 | 1.50 | 50 | 1.12 |
| L50 | {ped_injury=severe,lighting_condition=dark_with_streetlight,location_type=business_mixed_residential, road_type=two_no_separation} => {distance=D1} | 0.54 | 58.62 | 1.44 | 17 | 1.08 |
| L51 | {lighting_condition=dark_with_streetlight,location_type=business_mixed_residential,road_type=two_no_separation, ped_alcohol_drug=no} => {distance=D1} | 1.85 | 58.59 | 1.44 | 58 | 1.07 |
| L52 | {lighting_condition=dark_with_streetlight,location_type=business_mixed_residential,road_type=two_no_separation, posted_speed_limit=<30} => {distance=D1} | 0.57 | 58.06 | 1.43 | 18 | 1.07 |

## Table 12. Crash Pattern Identification by LIC for zone D2

| ID | rules | S(%) | C(%) | L | # | LIC |
|---|---|---|---|---|---|---|
| L53 | {primary_factor=driver_condition, driver_age=>64y} => {distance=D2} | 0.45 | 73.68 | 2.18 | 14 | N/A |
| L54 | {ped_injury=fatal,primary_factor=driver_condition,driver_age=>64y} => {distance=D2} | 0.10 | 100.00 | 2.96 | 3 | 1.36 |
| L55 | {primary_factor=driver_condition,driver_age=>64y,posted_speed_limit=50-55} => {distance=D2} | 0.10 | 100.00 | 2.96 | 3 | 1.36 |
| L56 | {crash_hour=12pm-6pm,primary_factor=driver_condition,driver_age=>64y} => {distance=D2} | 0.22 | 87.50 | 2.59 | 7 | 1.19 |
| L57 | {primary_factor=driver_condition,driver_age=>64y,driver_condition=inattentive_distracted} => {distance=D2} | 0.16 | 83.33 | 2.46 | 5 | 1.13 |
| L58 | {primary_factor=driver_condition,driver_age=>64y,driver_condition=illness_fatigued_asleep} => {distance=D2} | 0.13 | 80.00 | 2.37 | 4 | 1.09 |
| L59 | {primary_factor=driver_condition,driver_age=>64y,posted_speed_limit=<30} => {distance=D2} | 0.13 | 80.00 | 2.37 | 4 | 1.09 |
| L60 | {primary_factor=driver_condition,driver_age=>64y,vehicle_type=light_truck} => {distance=D2} | 0.22 | 77.78 | 2.30 | 7 | 1.06 |
| L61 | {crash_hour=12am-6am,primary_factor=prior_movement} => {distance=D2} | 0.67 | 53.85 | 1.59 | 21 | N/A |
| L62 | {crash_hour=12am-6am,primary_factor=prior_movement,location_type=residential} => {distance=D2} | 0.29 | 90.00 | 2.66 | 9 | 1.67 |
| L63 | {ped_injury=moderate,crash_hour=12am-6am,primary_factor=prior_movement} => {distance=D2} | 0.29 | 64.29 | 1.90 | 9 | 1.19 |
| L64 | {crash_hour=12am-6am,primary_factor=prior_movement,lighting_condition=dark_with_streetlight} => {distance=D2} | 0.57 | 62.07 | 1.84 | 18 | 1.15 |
| L65 | {crash_hour=12am-6am,primary_factor=prior_movement,road_type=two_no_separation} => {distance=D2} | 0.48 | 60.00 | 1.77 | 15 | 1.11 |
| L66 | {crash_hour=12am-6am,primary_factor=prior_movement,vehicle_type=passenger_car} => {distance=D2} | 0.32 | 58.82 | 1.74 | 10 | 1.09 |
| L67 | {crash_hour=12am-6am,primary_factor=prior_movement,driver_violation=no_violations} => {distance=D2} | 0.26 | 57.14 | 1.69 | 8 | 1.06 |
| L68 | {primary_factor=prior_movement,lighting_condition=dark_with_streetlight,location_type=residential} => {distance=D2} | 0.80 | 62.50 | 1.85 | 25 | N/A |
| L69 | {primary_factor=prior_movement,lighting_condition=dark_with_streetlight,location_type=residential,vehicle_type=light_truck} => {distance=D2} | 0.26 | 88.89 | 2.63 | 8 | 1.42 |
| L70 | {crash_hour=12am-6am,primary_factor=prior_movement,lighting_condition=dark_with_streetlight,location_type=residential} => {distance=D2} | 0.22 | 87.50 | 2.59 | 7 | 1.40 |
| L71 | {primary_factor=prior_movement,lighting_condition=dark_with_streetlight,location_type=residential,ped_age=41-64y} => {distance=D2} | 0.16 | 83.33 | 2.46 | 5 | 1.33 |
| L72 | {primary_factor=prior_movement,lighting_condition=dark_with_streetlight,location_type=residential,driver_age=41-64y} => {distance=D2} | 0.26 | 80.00 | 2.37 | 8 | 1.28 |
| L73 | {primary_factor=prior_movement,lighting_condition=dark_with_streetlight,location_type=residential,ped_age=25-40y} => {distance=D2} | 0.26 | 72.73 | 2.15 | 8 | 1.16 |
| L74 | {primary_factor=prior_movement,lighting_condition=dark_with_streetlight,location_type=residential,ped_age=15-24y} => {distance=D2} | 0.19 | 66.67 | 1.97 | 6 | 1.07 |

| L75 | {primary_factor=prior_movement,lighting_condition=dark_with_streetlight,location_type=residential,posted_speed_limit=<30} => {distance=D2} | 0.57 | 66.67 | 1.97 | 18 | 1.07 |
|---|---|---|---|---|---|---|
| L76 | {primary_factor=prior_movement,lighting_condition=dark_with_streetlight,location_type=residential,ped_condition=normal} => {distance=D2} | 0.32 | 66.67 | 1.97 | 10 | 1.07 |
| L77 | {road_type=two_no_separation,ped_alcohol_drug=yes,vehicle_type=Van_SUV} => {distance=D2} | 0.80 | 51.02 | 1.51 | 25 | N/A |
| L78 | {location_type=business_industrial,road_type=two_no_separation,ped_alcohol_drug=yes,vehicle_type=Van_SUV} => {distance=D2} | 0.26 | 66.67 | 1.97 | 8 | 1.31 |
| L79 | {lighting_condition=dark_no_streetlight,road_type=two_no_separation,ped_alcohol_drug=yes,vehicle_type=Van_SUV} => {distance=D2} | 0.32 | 62.50 | 1.85 | 10 | 1.23 |
| L80 | {primary_factor=pedestrian_actions,road_type=two_no_separation,ped_alcohol_drug=yes,vehicle_type=Van_SUV} => {distance=D2} | 0.51 | 61.54 | 1.82 | 16 | 1.21 |
| L81 | {road_type=two_no_separation,ped_alcohol_drug=yes,vehicle_type=Van_SUV,driver_violation=no_violations} => {distance=D2} | 0.73 | 60.53 | 1.79 | 23 | 1.19 |
| L82 | {crash_hour=6pm-12am,road_type=two_no_separation,ped_alcohol_drug=yes,vehicle_type=Van_SUV} => {distance=D2} | 0.54 | 56.67 | 1.68 | 17 | 1.11 |
| L83 | {lighting_condition=dark_no_streetlight,location_type=residential,posted_speed_limit=<30} => {distance=D2} | 0.89 | 50.00 | 1.48 | 28 | N/A |
| L84 | {crash_hour=12am-6am,lighting_condition=dark_no_streetlight,location_type=residential,posted_speed_limit=<30} => {distance=D2} | 0.16 | 83.33 | 2.46 | 5 | 1.67 |
| L85 | {ped_injury=moderate,lighting_condition=dark_no_streetlight,location_type=residential,posted_speed_limit=<30} => {distance=D2} | 0.41 | 65.00 | 1.92 | 13 | 1.30 |
| L86 | {lighting_condition=dark_no_streetlight,location_type=residential,posted_speed_limit=<30,vehicle_type=passenger_car} => {distance=D2} | 0.57 | 58.06 | 1.72 | 18 | 1.16 |
| L87 | {primary_factor=ped_violation,lighting_condition=dark_no_streetlight,location_type=residential,posted_speed_limit=<30} => {distance=D2} | 0.41 | 54.17 | 1.60 | 13 | 1.08 |

## Table 13. Crash Pattern Identification by LIC for zone D3

| ID | rules | S (%) | C (%) | L | # | LIC |
|---|---|---|---|---|---|---|
| L88 | {location_type=open_country} => {distance=D3} | 3.57 | 67.88 | 2.67 | 112 | N/A |
| L89 | {location_type=open_country,posted_speed_limit=>55} => {distance=D3} | 0.35 | 100.00 | 3.93 | 11 | 1.47 |
| L90 | {primary_factor=other_factors,location_type=open_country} => {distance=D3} | 0.54 | 94.44 | 3.71 | 17 | 1.39 |
| L91 | {location_type=open_country,vehicle_type=light_truck} => {distance=D3} | 1.28 | 78.43 | 3.08 | 40 | 1.16 |
| L92 | {crash_hour=6am-12pm,location_type=open_country} => {distance=D3} | 0.54 | 77.27 | 3.04 | 17 | 1.14 |
| L93 | {location_type=open_country,driver_condition=other_unknown} => {distance=D3} | 0.70 | 75.86 | 2.98 | 22 | 1.12 |
| L94 | {location_type=open_country,driver_age=unknown} => {distance=D3} | 0.48 | 75.00 | 2.95 | 15 | 1.10 |
| L95 | {location_type=open_country,ped_age=41-64y} => {distance=D3} | 1.12 | 74.47 | 2.93 | 35 | 1.10 |
| L96 | {location_type=open_country,driver_violation=others} => {distance=D3} | 0.93 | 72.50 | 2.85 | 29 | 1.07 |
| L97 | {lighting_condition=dark_no_streetlight,location_type=open_country} => {distance=D3} | 2.30 | 72.00 | 2.83 | 72 | 1.06 |
| L98 | {ped_injury=complaint,location_type=open_country} => {distance=D3} | 1.12 | 71.43 | 2.81 | 35 | 1.05 |
| L99 | {lighting_condition=dark_no_streetlight,posted_speed_limit=50-55} => {distance=D3} | 4.88 | 56.04 | 2.20 | 153 | N/A |
| L100 | {ped_injury=fatal,lighting_condition=dark_no_streetlight,posted_speed_limit=50-55} => {distance=D3} | 1.66 | 65.82 | 2.59 | 52 | 1.17 |
| L101 | {lighting_condition=dark_no_streetlight,posted_speed_limit=50-55,vehicle_type=light_truck} => {distance=D3} | 2.14 | 65.05 | 2.56 | 67 | 1.16 |
| L102 | {crash_hour=6am-12pm,lighting_condition=dark_no_streetlight,posted_speed_limit=50-55} => {distance=D3} | 0.83 | 65.00 | 2.55 | 26 | 1.16 |
| L103 | {lighting_condition=dark_no_streetlight,posted_speed_limit=50-55,driver_violation=careless_operation} => {distance=D3} | 0.35 | 61.11 | 2.40 | 11 | 1.09 |
| L104 | {primary_factor=pedestrian_actions,lighting_condition=dark_no_streetlight,posted_speed_limit=50-55} => {distance=D3} | 2.26 | 59.66 | 2.34 | 71 | 1.06 |
| L105 | {road_type=two_no_separation,posted_speed_limit=50-55} => {distance=D3} | 6.54 | 50.87 | 2.00 | 205 | N/A |
| L106 | {location_type=open_country,road_type=two_no_separation,posted_speed_limit=50-55} => {distance=D3} | 2.04 | 71.11 | 2.79 | 64 | 1.40 |
| L107 | {ped_injury=fatal,road_type=two_no_separation,posted_speed_limit=50-55} => {distance=D3} | 1.69 | 62.35 | 2.45 | 53 | 1.23 |
| L108 | {primary_factor=pedestrian_condition,road_type=two_no_separation,posted_speed_limit=50-55} => {distance=D3} | 0.32 | 58.82 | 2.31 | 10 | 1.16 |
| L109 | {primary_factor=other_factors,road_type=two_no_separation,posted_speed_limit=50-55} => {distance=D3} | 0.73 | 56.10 | 2.20 | 23 | 1.10 |
| L110 | {lighting_condition=dark_no_streetlight,road_type=two_no_separation,posted_speed_limit=50-55} => {distance=D3} | 4.18 | 55.74 | 2.19 | 131 | 1.10 |
| L111 | {road_type=two_no_separation,posted_speed_limit=50-55,vehicle_type=light_truck} => {distance=D3} | 2.62 | 55.41 | 2.18 | 82 | 1.09 |
| L112 | {location_type=business_industrial,road_type=two_no_separation,posted_speed_limit=50-55} => {distance=D3} | 0.38 | 54.55 | 2.14 | 12 | 1.07 |
| L113 | {crash_hour=6am-12pm,road_type=two_no_separation,posted_speed_limit=50-55} => {distance=D3} | 1.18 | 53.62 | 2.11 | 37 | 1.05 |
| L114 | {posted_speed_limit=50-55,driver_violation=no_violations} => {distance=D3} | 4.94 | 50.00 | 1.96 | 155 | N/A |
| L115 | {location_type=open_country,posted_speed_limit=50-55,driver_violation=no_violations} => {distance=D3} | 1.47 | 75.41 | 2.96 | 46 | 1.51 |
| L116 | {ped_condition=normal,posted_speed_limit=50-55,driver_violation=no_violations} => {distance=D3} | 0.70 | 59.46 | 2.34 | 22 | 1.19 |
| L117 | {ped_injury=fatal,posted_speed_limit=50-55,driver_violation=no_violations} => {distance=D3} | 1.63 | 57.30 | 2.25 | 51 | 1.15 |
| L118 | {ped_age=41-64y,posted_speed_limit=50-55,driver_violation=no_violations} => {distance=D3} | 1.75 | 57.29 | 2.25 | 55 | 1.15 |

| L119 | {lighting_condition=dark_no_streetlight,posted_speed_limit=50-55,driver_violation=no_violations} => {distance=D3} | 3.35 | 57.07 | 2.24 | 105 | 1.14 |
|---|---|---|---|---|---|---|
| L120 | {driver_age=25-40y,posted_speed_limit=50-55,driver_violation=no_violations} => {distance=D3} | 1.72 | 53.47 | 2.10 | 54 | 1.07 |
| L121 | {crash_hour=6am-12pm,posted_speed_limit=50-55,driver_violation=no_violations} => {distance=D3} | 0.99 | 53.45 | 2.10 | 31 | 1.07 |
| L122 | {crash_hour=12pm-6pm,posted_speed_limit=50-55,driver_violation=no_violations} => {distance=D3} | 0.54 | 53.13 | 2.09 | 17 | 1.06 |
| L123 | {ped_alcohol_drug=no,posted_speed_limit=50-55,driver_violation=no_violations} => {distance=D3} | 2.68 | 52.83 | 2.08 | 84 | 1.06 |
| L124 | {ped_age=25-40y,posted_speed_limit=50-55,driver_violation=no_violations} => {distance=D3} | 1.88 | 52.68 | 2.07 | 59 | 1.05 |